# Performance of graphene Hall effect sensors: role of bias current, disorder and Fermi velocity

**Lionel Petit[1], Tom Fournier[1,2], Géraldine Ballon[1], Cédric Robert[1], Delphine Lagarde[1], Pascal Puech[2], Thomas Blon[1] and Benjamin Lassagne[1]**

[1] Université de Toulouse, INSA-CNRS-UPS, LPCNO, 135 Av. Rangueil, 31077 Toulouse, France
[2] Centre d'Elaboration des Matériaux et d'Etudes Structurales (CEMES), UPR8011 CNRS, Université Toulouse 3, 31055 Toulouse, France

E-mail: lassagne@insa-toulouse.fr

Graphene-based Hall effect magnetic field sensors hold great promise for the development of ultrasensitive magnetometers with very low power consumption. Their performance is frequently analyzed using the so-called two-channel model where electron and hole conductivities are simply added. Unfortunately, this model is unable to capture all the sensor's characteristics, particularly the bias current dependence of the magnetic field sensitivity. Here, we present an advanced model that provides an in-depth understanding of how graphene-based Hall sensors operate and demonstrate its ability to quantitatively assess their performance. First, we report on the fabrication of sensors based on different qualities of graphene, with the best devices achieving magnetic field sensitivities as high as $5000\ \Omega/T$, outperforming the best silicon and narrow-gap semiconductor-based sensors. Then, we examine their performance in detail using the proposed numerical model, which combines Boltzmann formalism with distinct Fermi levels for electrons and holes, and a new method for the introduction of substrate-induced electron–hole puddles. Importantly, the dependences of magnetic field sensitivity on bias current, disorder, substrate, and Hall bar geometry are quantitatively reproduced for the first time. In addition, the model emphasizes that the performance of devices with widths of the order of the charge carrier diffusion length is significantly affected by the bias electric current due to the occurrence of charge carrier accumulation and depletion areas near the edges of the Hall bar, much larger than conventional Hall effect predictions. The formation of these areas induces a transverse diffusion charge carrier flux capable of counterbalancing the one induced by the Lorentz force when the Hall electric field cancels out in the ambipolar regime. Finally, we discuss how sensor performance can be enhanced by Fermi velocity engineering, paving the way for future ultrasensitive graphene-based Hall effect sensors.

## I. INTRODUCTION

The market for magnetic field sensors is currently very large, representing about 3 billion euros in 2023, with applications in a wide range of fields such as automotive, consumer electronics, position and motion sensing, magnetic storage, magnetic field mapping, and biosensing. It also concerns fundamental research on magnetic [1] or superconducting [2] materials. To date, the most utilized magnetic field sensor technology is based on the measurement of the well-known transverse Hall voltage $U_h$, which appears in a thin, long, bar-shaped conductive material supplied with a bias current $I$ and immersed in a perpendicular magnetic field $B$ [3]. The success of Hall effect sensors is based on this very simple and nonperturbative measurement scheme, combined with a linear response over a wide range of magnetic fields and temperatures [3–5]. Such versatility is not accessible with superconducting quantum interference devices [6] or magnetoresistive sensors based on giant or tunnel magnetoresistance effects [6], which are limited to cryogenic temperatures and low magnetic fields, respectively.

New directions for the development of Hall sensors were made possible by the first isolation of graphene in 2004 [7]. With an intrinsic one-atom thick structure, very high charge carrier mobility [8] and low charge carrier density, graphene Hall sensors (GHSs) show an unprecedented magnetic field sensitivity ($S_I = U_h/(B \times I)$) and outperform state-of-the-art silicon and narrow-gap semiconductor Hall sensors [4,6,9–13]. As a result, GHSs have been demonstrated as either ultrasensitive magnetic field sensors for magnetometry [14,15] and nanoscale magnetic field mapping [13] or sensors having Hall voltages of the same order of magnitude as those obtained with silicon sensors but with 100 times lower power consumption [6]. This latter point represents a significant advance and paves the way for the development of ultralow power devices.

The trend toward ever-improving sensor performance requires a thorough understanding of the physical principles that govern their operation. In the case of GHSs, the Hall effect is frequently analyzed through the so-called two-channel model [4,11,16,17], which intrinsically possesses limitations in the capture of all the GHSs reported features. In particular, the bias current dependence of $S_I$ observed in several previous works [6,11,17–19] cannot be explained since the two-channel model equations were obtained for an electrodeless Hall bar with uniform charge carrier doping and infinite length. In this paper, we analyze the Hall effect in GHSs of different quality and size fabricated with graphene grown by chemical vapor deposition (CVD) or graphene exfoliated from a crystal of highly orientated pyrolytic graphite (HOPG) using a more advanced physical model combining several approaches. The charge carrier transport properties are described through the Boltzmann formalism, where electrons and holes are treated with distinct Fermi levels in contrast to the two-channel model. A local field effect model is used to account for the effect of gate voltage and bias current on electron and hole doping, and a new semiempirical model of electron and hole puddles is developed to account for the effect of impurities. As a result, we can quantitatively describe all the galvanomagnetic features of GHSs for the first time. In particular, we highlight and explain how the bias current affects the spatial profile of the electron and hole doping inside the GHSs, resulting in a significant modification of the shape of $S_I$ as a function of gate voltage and bias current, especially for devices having a width comparable to the charge carrier diffusion length. This later point, which is a consequence of the ambipolar nature of graphene, has never been addressed before. Also, we shed light on the role of electron–hole puddles and substrates on $S_I$, thus revealing a new way to improve GHS performance by engineering the Fermi velocity of graphene.

## II. EXPERIMENTAL RESULTS
### A. Electrical characteristics of Hall bars

GHSs were fabricated either from CVD graphene monolayers (from Graphenea) transferred using a semidry technique onto $90 \pm 10$ nm-thick SiO$_2$/Si-doped substrates (referred to as CVD-GHSs in the following) or from HOPG graphene monolayers (from HQ Graphene) deposited by mechanical exfoliation onto $285 \pm 10$ nm-thick SiO$_2$/Si heavily-doped substrates (referred to as HOPG-GHSs in the following). A third kind of GHSs were fabricated by encapsulation of a HOPG graphene monolayer in hexagonal boron nitride (h-BN, from HQ Graphene) using a pick-up technique [20] and which was subsequently transferred onto $285 \pm 10$ nm-thick SiO$_2$/Si heavily-doped substrates (referred to as hBN-GHSs in the following). Polymethylmethacrylate masks, typically 8-branch Hall bars (defined through electron-beam lithography), were used to etch the graphene with successive SF$_6$ and O$_2$ plasma, the former being used only for hBN-GHSs. Electrodes were then patterned by electron-beam lithography followed by Cr (5 nm) then Au (25 nm for HOPG and hBN-GHSs and 200 nm for CVD-GHSs) Joule heating evaporation. The typical GHS width and length dimensions are 5 and 60 μm, 1 and 17 μm, and 2 and 11 μm for CVD, HOPG, and hBN-GHSs, respectively. Finally, electrical measurements were performed at low He pressure under a magnetic field at 0.1 and 1 T. CVD-GHSs were measured at 200 K to avoid hysteresis on the gate voltage while HOPG and hBN-GHSs, for which there is no hysteresis, were measured at 300 K.

Several samples of each kind (i.e., CVD-GHSs, HOPGGHSs, and hBN-GHSs) were fabricated, and their characteristics were highly reproducible. Figures 1(a)–1(f) show typical Raman spectra and optical images of our devices. The Raman spectra [Figs. 1(a)–1(c)] display G ($\approx 1580 \ cm^{-1}$) and 2D peaks ($\approx 2680 \ cm^{-1}$) with ratios $I_{2D}/I_G \approx 2.41$, $I_{2D}/I_G \approx 2.82$ and $I_{2D}/I_G \approx 3.9$ for CVD-GHSs, HOPG-GHSs, and hBN-GHSs, respectively, confirming that our devices are made of monolayer graphene [21]. Note that the presence of the D ($1342 \ cm^{-1}$) and D′ ($1623 \ cm^{-1}$) peaks in Fig. 1(a) is characteristic of monolayers containing impurities and grain boundaries obtained by CVD. Figures 1(g)–1(l) show typical compensated longitudinal resistance $R_{LC} = \left(R_L(B) + R_L(-B)\right)/2$ and typical compensated magnetic field sensitivity $S_{IC} = (S_I(B) + S_I(-B))/2$ for different bias current values. We observe the expected shapes for $R_{LC}$ and $S_{IC}$ [4,6,10–12] with a maximum $S_{IC}$ reaching 5000 Ω/T for hBN-GHSs [black line on Fig. 1(l)], a value identical to the best report to date on similar structures [10]. Moreover, on Figs. 1(j)–1(l), we observe a pronounced effect of the bias current on the shape of $S_{IC}$, as previously reported in similar devices [6,11,17–19]. The gate voltage position of the charge neutrality point (CNP), where $S_{IC}$ inverts, shifts toward positive voltages as the current increases that results, in the case of CVD-GHSs, in a quasirigid translation of $S_{IC}$ as a function of $V_g$. For HOPG and hBN-GHSs, the increase in bias current also leads to a CNP shift but is accompanied by a drop of the maximum $|S_{IC}|$ as well as an increase in the distance separating $S_{IC}$ extrema. This effect is very significant for hBN-GHSs since it results in a large shape $S_{IC}$ modification, as previously reported by Shaeffer *et al*. [6].

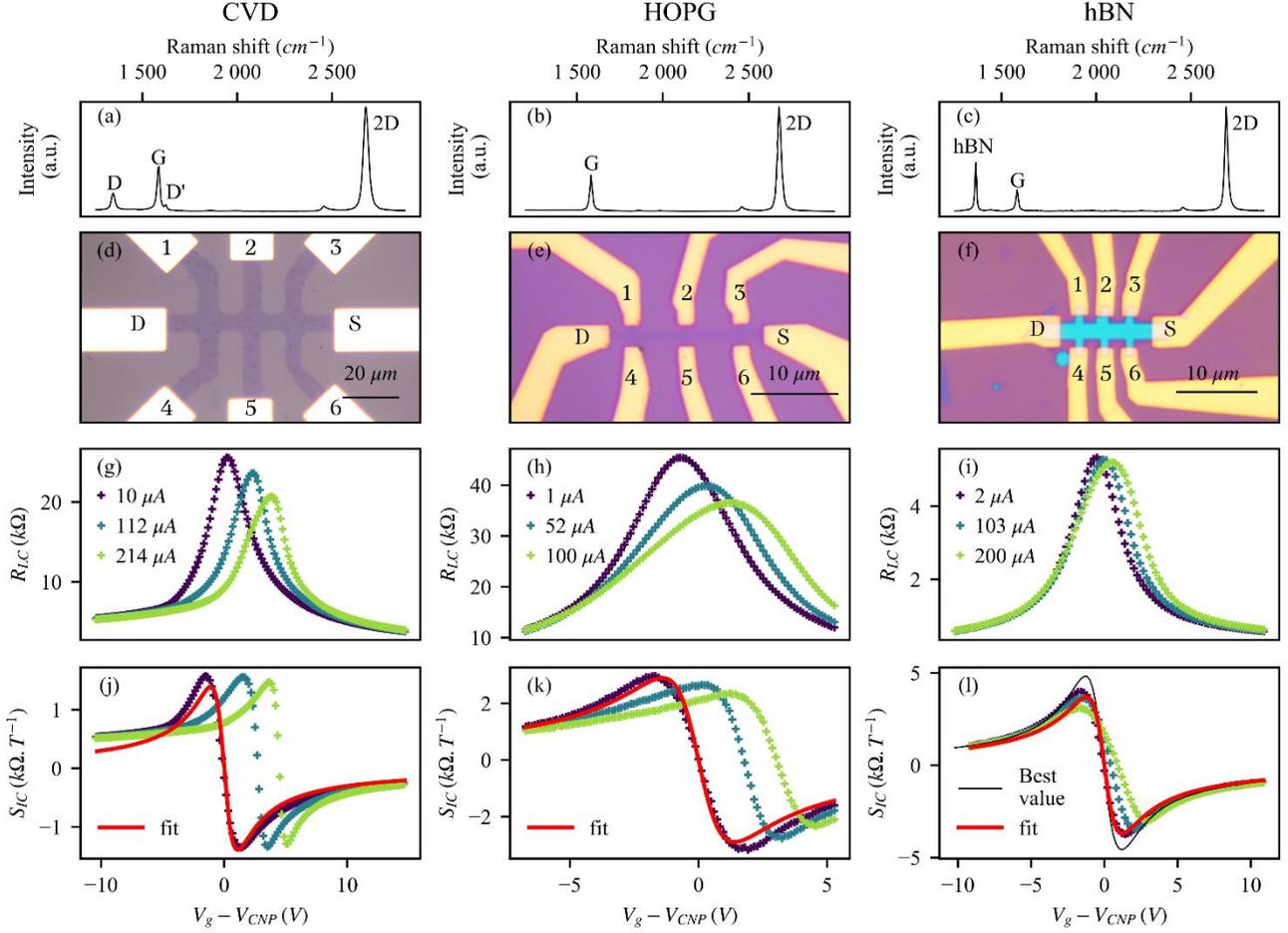

FIG. 1. Typical Raman spectra of CVD-GHS (a), HOPG-GHS (b) and hBN-GHS (c). Typical optical images of CVD-GHS (d), HOPG-GHS (e) and hBN-GHS (f). (g), (h) and (i) Typical compensated longitudinal 4 probe resistance $R_{LC}$ measured between contacts 1 and 3 as a function of gate voltage $V_g$ and for different bias current values. ((g) 200 K, 1 T, $V_{CNP} = 5.32\ V$ ; (h) 300 K, 0.1 T, $V_{CNP} = 7.73\ V$ ; (i) 300 K, 0.1 T, $V_{CNP} = -2.88\ V$). (j), (k) and (l) Typical compensated magnetic field sensitivity $S_{IC}$ as a function of $V_g$ for different bias current values (measured between contacts 1 and 4 for (j) and (k) and 2 and 5 for (l), same $T$ and $B$ values as for $R_{LC}$). The red line curves are the fit of the low-bias current $S_{IC}$ using Eq. 1 and 2 and obtained with the following fit parameters: (j) $t_{ox} = 105\ nm$, $n_0 = 2.24 \times 10^{11}\ cm^{-2}$, (k) $t_{ox} = 280\ nm$ and $n_0 = 1.08 \times 10^{11}\ cm^{-2}$ and (l) $t_{ox} = 310\ nm$ and $n_0 = 8.5 \times 10^{10}\ cm^{-2}$.

## B. Analysis with the two-channel model

In a first step, we analysed $S_I$ at low bias current using the two-channel model, as previously reported [4,11,16,17]. This model provides a fairly simple expression for $S_I$, when the charge carrier mobility $\mu_n$ and $\mu_p$ are equal and $\mu_{n(p)}B \ll 1$, which is written as follows

$$S_I = -\frac{1}{e}\frac{(n-p)}{(n+p)^2} \qquad (1)$$

where $e$ is the elementary charge; $n$ and $p$ are the electron and hole densities, respectively. The gate voltage dependence of $S_I$ is taken into account using empirical relationships established by Meric *et al.* [22] so that

$$n + p \approx \sqrt{n_0^2 + n(V_g)^2} \quad (2a)$$

$$n - p = n(V_g) = \frac{C_g}{e}(V_g - V_{CNP}) \quad (2b)$$

where $n_0$ is the minimal charge carrier doping determined by temperature and residual impurities, $C_g$ is the gate capacitance per surface unit, $V_g$ is the gate voltage and $V_{CNP}$ is the gate voltage at CNP. The red line curves on the Figs. 1(j)-(l) are the fits of the experimental data using Eqs. 1 and 2. The oxide thicknesses $t_{ox}$ were determined from the gate capacitance values and equal to $105\ nm$, $280\ nm$, and $310\ nm$ for the CVD, HOPG and hBN-GHS respectively. Regarding the residual doping $n_0$, we obtained values of $2.24 \times 10^{11} cm^{-2}$, $1.08 \times 10^{11} cm^{-2}$ and $8.5 \times 10^{10} cm^{-2}$ for CVD, HOPG, and hBN-GHSs respectively, which is consistent with the fact that CVD graphene monolayers have more defects and the hBN encapsulation protects graphene from charged impurities and process contamination. Equations 1 and 2 are quite convenient for an initial analysis of $S_I$ at a low bias current, as they allow for both an understanding of the effect of the doping on the GHS sensitivity and extract important parameters such as the residual doping value or the gate capacitance. Unfortunately, this approach remains limited, particularly for anticipating the device geometry and bias current effects on $S_I$. Therefore, further developments require an advanced model that should consider the exact GHS geometry as well as the influence of the gate voltage and bias current on the charge carrier doping spatial distribution.

## III. ADVANCED MODEL

The model we developed is based on the combination of several approaches used to describe the graphene transport properties in the diffusive regime. Thermal charge carrier doping and transport properties are described using the Boltzmann formalism and electrostatic doping is taken into account thanks to a field effect model, which has been undertaken previously for graphene field effect transistors [22–25]. In the present work, some refinements have been made. Distinct electron and hole Fermi levels have been used and recombination-generation processes have been added to describe the operation of the GHSs in the ambipolar regime more precisely. We have also developed a new method to account for the presence of impurities.

### A. Charge carrier thermal statistics and electric charge

The temperature dependence of the electron and hole doping $n$ and $p$ is taken into account by the following expressions:

$$n(E_{fn}) = \int_{E_{CNP}}^{\infty} f_n(E) D(E) dE \quad (3a)$$

$$p(E_{fp}) = \int_{-\infty}^{E_{CNP}} \left(1 - f_p(E)\right) D(E) dE \quad (3b)$$

where $D(E) = 2|E - E_{CNP}|/\pi(\hbar v_f)^2$ is the graphene density of states, $E$ is the electron energy for a state with a wave vector $\vec{k}$, $E_{CNP}$ is the energy at the CNP, $\hbar$ is the reduced Plank constant, and $v_f$ is the Fermi velocity. Moreover, $f_n(E)$ and $f_p(E)$ are the Fermi-Dirac distribution of the electrons and holes, respectively, given by $f_{n(p)}(E) = 1/\left(1 + exp\left((E - E_{fn(p)})/k_B T\right)\right)$ with $k_B$ as the Boltzmann's constant, $T$ the temperature, and $E_{fn(p)}$ the electron (hole) Fermi level. We also define the shift of the Fermi level $\Delta E_{fn(p)}$ as $\Delta E_{fn(p)} = E_{fn(p)} - E_{CNP}$ [Fig. 2(b)].

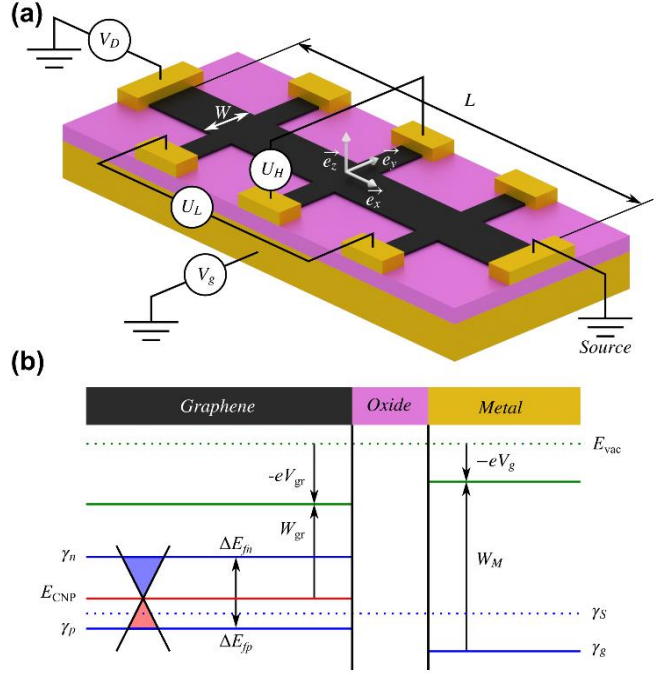

FIG. 2. (a) Typical device geometry. $V_D$ is the drain electrostatic potential, the source is the reference electrostatic potential set to zero. (b) Band diagram of the graphene and gate electrode. $E_{vac}$ corresponds to the electron energy in vacuum.

Regarding the effect of the gate voltage and bias current on the electron and hole doping in graphene, we used a field effect model [22–25]. In this framework, the source is used as the reference electrostatic potential. Its Fermi level represents the origin of the energy; thus, its electrochemical potential $\gamma_S$ is equal to $0\ eV$ and the electrochemical potential $\gamma_D$ of the drain electrode, with tunable electrostatic potential $V_D$, is equal to $-eV_D$ [Fig. 2(a)]. For the sake of simplicity, the doped silicon substrate that serves as the back-gate electrode is modeled by a metallic electrode [Fig. 2(a)] having an electrochemical potential $\gamma_g$ equal to $-eV_g$ [Fig. 2(b)]. Finally, the electrochemical potentials $\gamma_n$ and $\gamma_p$ of the electrons and holes in graphene are defined by $\gamma_{n(p)} = E_{fn(p)} - eV_{gr}$, with $V_{gr}$ as the electrostatic potential in graphene [Fig. 2(b)]. The graphene CNP energy [red line in Fig. 2(b)] is defined as $E_{CNP} = W_M - W_{gr} - eV_{gr}$, with $W_M$ and $W_{gr}$ as the metal and graphene work function, respectively. Then, using Gauss's law [24–26], a simple and local relationship between $V_g$, $V_{gr}$ and the charge carrier density in the graphene sheet $Q_{gr} = -e\big(n(\gamma_n + eV_{gr}) - p(\gamma_p + eV_{gr})\big)$ can be found so that

$$Q_{gr}(x,y) + Q_g(x,y) + Q_0(x,y) = 0 \qquad (4)$$

where $Q_g = C_g(V_g - V_{gr})$ is the local electric charge density in the gate electrode and $Q_0$ is a possible residual electric charge density coming from contamination, impurities or work function differences.

**B. Transport properties**

To describe the galvanomagnetic properties of GHSs in a perpendicular magnetic field [along $z$, Fig. 2(a)], we used the Boltzmann formalism [3,23]. The electric current density $\vec{J}$ in graphene is the sum of the contribution of the electrons, $\vec{J}_n$, and holes, $\vec{J}_p$, i.e., $\vec{J} = \vec{J}_n + \vec{J}_p$ with

$$\vec{J}_{n(p)} = \bar{\bar{\sigma}}_{n(p)} \frac{\vec{\nabla}\gamma_{n(p)}}{e} \qquad (5)$$

$\bar{\bar{\sigma}}_{n(p)}$ is the antisymmetric conductivity tensor for electrons (holes). Its components $\sigma_{xx_{n(p)}}$ and $\sigma_{xy_{n(p)}}$ are given by,

$$\sigma_{xx_{n(p)}} = \frac{\eta e}{\pi v_f^2 \hbar^2} \int_{E_{CNP}}^{-\eta \times \infty} \frac{(E-E_{CNP})^2 \mu(E)}{1+\mu^2(E)B^2} \frac{\partial f_{n(p)}}{\partial E} dE \quad (6a)$$

$$\sigma_{xy_{n(p)}} = \frac{eB}{\pi v_f^2 \hbar^2} \int_{E_{CNP}}^{-\eta \times \infty} \frac{(E-E_{CNP})^2 \mu^2(E)}{1+\mu^2(E)B^2} \frac{\partial f_{n(p)}}{\partial E} dE \quad (6b)$$

where $\eta = -1$ for electrons and $\eta = 1$ for holes. The charge carrier mobility $\mu(E)$ depends on the charge carrier scattering mechanisms (i.e., charged impurities, vacancies, ripples, phonons…) [26,27] and is expressed in terms of the scattering time $\tau_i$ of each mechanism by $\mu(E) = ev_f^2 |E - E_{CNP}|^{-1}(\sum_i 1/\tau_i(E))^{-1}$. Equations 5 and 6, where the electrochemical potentials are considered unequal for electrons and holes, allow for both conduction and diffusion currents. The latter can have a significant contribution when a p-n junction forms in the GHS [23] or when charge carriers accumulate and deplete near the edges of the Hall bar, especially in the ambipolar regime [28] (see Sec. IV).

In this work we focus on the stationary regime, hence conservation of electric charge implies that the divergences of electron and hole current densities are given by

$$\vec{\nabla}.\vec{J}_{n(p)} = -\eta eR = -\eta ek(np - n_{eq}p_{eq}) \quad (7)$$

Here, $R$ is the recombination-generation rate of the charge carriers. Different processes can participate in the electron-hole pair recombination-generation such as Auger scattering, optical and acoustic phonon scattering, or impurity assisted scattering [29,30] but the use of the exact $R$ expression is beyond the scope of this article. Instead, we use a linear expression (see Eq. 7) characterized by a constant $k$, specific of the recombination-generation process and by $n_{eq}$ and $p_{eq}$ the electron and hole doping at the equilibrium. This latter expression gives a fairly good account of the dependence of the recombination-generation rate on electron and hole doping for a small deviation from equilibrium. Finally, $\gamma_n$, $\gamma_p$ and $V_{gr}$, the solutions of the coupled Eqs. 4, 5 and 7, are obtained using the finite-element method with appropriate geometry and boundary conditions.

**C. Electron-hole puddle model**

Electron-hole puddles in graphene have been observed experimentally via scanning tunnelling microscopy (STM) [31,32]. They originate from charged impurities in the substrate and/or fabrication process contamination [26], or from graphene sheet deformation [33]. These electron-hole puddles play an important role in the electronic transport properties of graphene, introducing long-range type charge carrier scattering and residual charge carrier doping, which lead to a non-universal minimum conductivity [26,27,34].

To account for this effect, we developed a semi-empirical method to generate maps of random electron-hole puddles that was guided by experimental observations [31,32,35], theoretical calculations based on the Thomas-Fermi-Dirac formalism [36] and empirical modelling [22,37]. This method consists of generating maps of random Fermi level fluctuations $\Delta E_{fr}$ by adding a given density $n_{0l}$ of Lorentzian like function $L_i(x,y) = \left(1 + \|\overrightarrow{O_iM}\|^2 / r_{0i}^2\right)^{-1/2}$ with $x$, $y$, the $M$ coordinates, $r_{0i}$ as a random radius and $O_i(x_{oi}, y_{oi})$, as a random position. The map of $\Delta E_{fr}$ was calculated using the following ansatz

$$\Delta E_{fr}(x,y) = \Delta E_{frmax} \tanh\left(\sum_{i=1}^{n_{0l} \times A} L_i(x,y)\right) \quad (8)$$

where $\Delta E_{frmax}$ is the maximum amplitude of the fluctuations of the Fermi level and $A$ is the area of the sensor. As explained by Rossi *et al*. [36], who developed an effective medium theory to study the electronic transport in disordered graphene, the puddles that primarily contribute to the charge carrier transport concern wide regions spanning the system size, presenting a low charge carrier density and an almost uniform conductivity. Equation (8) is perfectly adapted to generate such wide charge carrier puddles [Fig. 10(a)] and to reproduce qualitatively Fermi-level fluctuations observed experimentally [31,32,35]. Finally, to include the effect of electron–hole puddles in the previously presented model, the map of random Fermi-level fluctuations is converted into a map of random gate voltage fluctuations $\Delta V_g(x,y)$ [see Sec. I, Eq. (s1) of the Supplemental Material [38]] and added to Eq. (4), through the electric charge density in the gate, as follows: $Q_g(x,y) = C_g(V_g + \Delta V_g(x,y) - V_{gr}(x,y))$.

## IV. RESULTS OF THE ADVANCED MODEL

In this section, we report on the main findings of the physical model presented in Sec. III. First, an in-depth comparison with the two-channel model highlights what new physical mechanisms our advanced model brings. Then, a focus is made on how the puddles, bias current, and geometry affect $R_L$ and $S_I$. For all the explanations given in the following, the simulations were performed on 8-branch Hall bars of width $W$, electrode width $W_h = W/i_h$, total transversal length $W' = i' \times W$ with $i_h$ and $i'$ as positive integers, and total longitudinal length $L = 8W + 3W_h$ [Fig. 2(a)], which allows the aspect ratio and resistance to remain constant as $W$ varies. Each branch of the Hall bar was connected to a metallic electrode of length $W/2$ with a constant conductivity $\sigma_c = 1\,S$ corresponding to a 25-nm-thick gold film. For the sake of simplicity, $W_M$ and $W_{gr}$ were considered equal and $Q_0$ was set to zero, the main effect of these parameters being to shift the CNP gate position. We fixed $\gamma_n = \gamma_p = \gamma_D$ at the drain electrode and $\gamma_n = \gamma_p = 0$ at the source electrode. Moreover, we imposed $\gamma_n = \gamma_p$ on the boundaries between graphene and metallic electrodes in order to keep them equal inside the electrodes. The Hall voltage was defined as $-eU_h = \gamma_{n(p)}(0, -W'/2) - \gamma_{n(p)}(0, W'/2)$ and the longitudinal four-probe voltage as $-eU_L = \gamma_{n(p)}(-2W - W_h, W'/2) - \gamma_{n(p)}(2W + W_h, W'/2)$ [Fig. 2(a)]. In the simulations presented below, we used the following parameters: $v_f = 10^6\,m/s$, $B = 100\,mT$, $T = 300\,K$ and $C_g = 1.15\,F/m^2$ corresponding to an oxide thickness $t_{ox}$ of 300 nm. Regarding $k$, we used a value equal to $10^{-4}\,m^2/s$ corresponding to recombination times $\tau_r \approx 1/k(n_{eq} + p_{eq})$ ranging from 1 ps to 10 ps depending on the doping [see Sec. II, Eq. (s7) of the Supplemental Material [38]]. These values agreed with the theoretical predictions at room temperature for pristine graphene [29,30]. The scattering of charge carriers was considered to originate from long range disorder, which means the charge carrier mobility $\mu$ is constant and equal for electrons and holes [26,27]. Regarding $n_{eq}$ and $p_{eq}$, they were determined considering that electrons and holes have the same Fermi level $E_{feq}$ and that the equilibrium electric charge $Q_{eq}(E_{feq}) = -e(n_{eq}(E_{feq}) - p_{eq}(E_{feq}))$ is equal to $Q_{gr}(E_{fn}, E_{fp})$.

### A. Comparison with the two-channel model

A major difference between our advanced model and the two-channel model is that the electron and hole currents are calculated using different electrochemical potentials in the former case and only the electrostatic potential for the latter, meaning that the electric current densities are simply written in the two-channel model as $\vec{J}_{n(p)} = -\bar{\bar{\sigma}}_{n(p)} \vec{\nabla} V_{gr}$. To understand why the two-channel model is not sufficient to correctly capture the operation of a GHS, we need to consider not only the electric charge and associated electric current densities but also the charge carrier densities and associated fluxes $\vec{P}_{n(p)} = \eta \vec{J}_{n(p)}/e$. At a low bias current and a low magnetic field, the total transverse electric current density $J_y$ and total charge carrier flux $P_y$ in the two-channel model are written as [see Sec. II, Eqs. (s2) and (s3) of the Supplemental Material [38]]

$$J_y = -e(n-p)\mu^2 B \frac{\partial V_{gr}}{\partial x} - e(n+p)\mu \frac{\partial V_{gr}}{\partial y} \quad (9a)$$

$$P_y = (n+p)\mu^2 B \frac{\partial V_{gr}}{\partial x} + (n-p)\mu \frac{\partial V_{gr}}{\partial y} \quad (9b)$$

Equations (9a) and (9b) reveal why the two-channel model is inconsistent when describing the GHS operation, in particular, at the CNP when $n = p$ and the electric charge on the Hall bar edges is null. Indeed, even if $J_y$ is nullified by the removal of the transverse electric field $\partial V_{gr}/\partial y$ due to the absence of electric charge and the equality of $n$ and $p$, $P_y$ is not and is written as $P_y = 2n\mu^2 B \times \partial V_{gr}/\partial x$. It is inconsistent for a finite structure. This contribution to the charge carrier flux is caused by the Lorentz force, which deflects both types of charge carriers in the same transverse direction [28], normally leading to an excess of charge carriers near one edge and a deficit on the opposite one, consistent with a zero electric charge. This issue is not taken into account by the two-channel model, which is focused on the electric charge and not the charge carrier densities. It can be solved by the introduction of different electrochemical potentials for electrons and holes, allowing for their accumulation or depletion near the edges. In our advanced model, the total transverse electric current density $J_y$ and total charge carrier flux $P_y$ are written as follows [see Sec. II, Eqs. (s4) and (s5) of the Supplemental Material [38]]:

$$J_y = -e(n-p)\mu^2 B \frac{\partial V_{gr}}{\partial x} + \mu\left(n\frac{\partial E_{fn}}{\partial y} + p\frac{\partial E_{fp}}{\partial y}\right) - e(n+p)\mu\frac{\partial V_{gr}}{\partial y} \quad (10a)$$

$$P_y = (n+p)\mu^2 B \frac{\partial V_{gr}}{\partial x} - \frac{\mu}{e}\left(n\frac{\partial E_{fn}}{\partial y} - p\frac{\partial E_{fp}}{\partial y}\right) + (n-p)\mu\frac{\partial V_{gr}}{\partial y} \quad (10b)$$

As seen in Eq. (9b), the first term of Eq. (10b) is caused by the Lorentz force (noted as $P_{yL}$ in the following), while the second term, which depends on the transverse gradient of the Fermi levels ($\partial E_{fn}/\partial y$, $\partial E_{fp}/\partial y$), is a transverse diffusion flux induced by the accumulation and depletion of charge carriers near the edges of the Hall bar (noted as $P_{yD}$ in the following) and the third term is simply the Hall electric field term (noted as $P_{yH}$ in the following). At the CNP (where $n = p$), as seen previously, $P_{yH}$ cancels and only $P_{yD}$ can counterbalance $P_{yL}$, solving the inconsistency of the two-channel model. We can also note that when $n = p$, $\partial E_{fp}/\partial y = -\partial E_{fn}/\partial y$, which ensures that the electric current density also cancels.

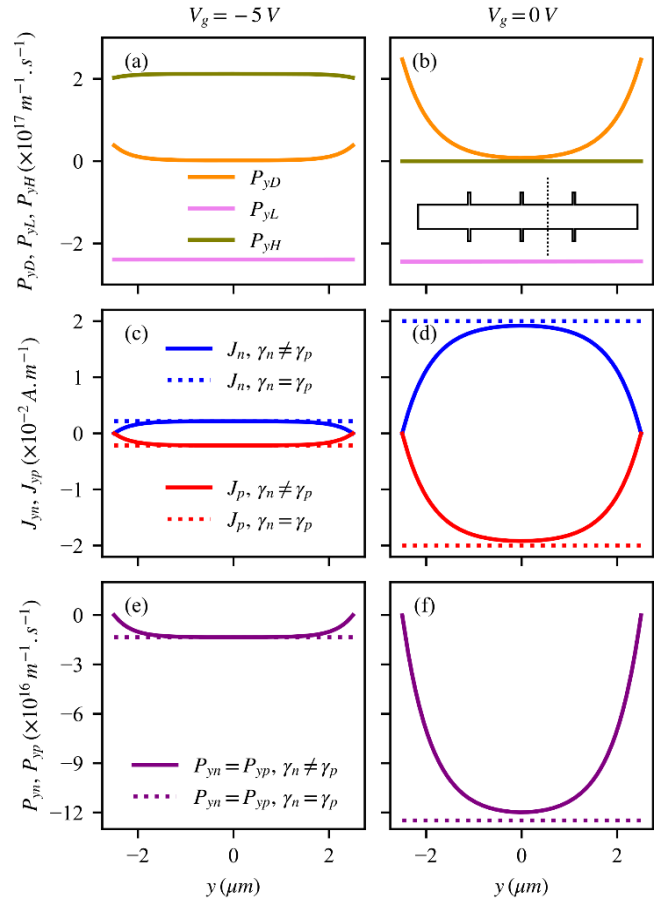

FIG. 3. (a) and (b) Different contributions of $P_y$ for two $V_g$ along one transverse cut of the Hall bar (see insert in (b)). $P_{yL}$ is the Lorentz component, $P_{yD}$, the diffusion component and $P_{yH}$ the Hall component. Electron and hole current densities (c) and (d) and charge carrier fluxes (e) and (f) for two $V_g$ calculated with our model (solid line) and in the case $\gamma_n = \gamma_p = \gamma$ (dotted lines).

To illustrate the above discussion, simulations were performed at a low bias current (i.e., 1 µA) on a Hall bar having a $\mu = 2\ m^2/(V.s)$ and a width $W = 5\ \mu m$. We used $W' = 2W$ and $W_h = W/10$ to ease the comparison with the two-channel model, which was established for an electrodeless Hall bar. Figures 3(a) and 3(b) show the different contributions of $P_y$ for two gate voltages ($V_g = -5\ V$ and $V_g = 0\ V$) along one transverse cut of the Hall bar [see the insert in Fig. 3(b)]. We can note that the shape of the different contributions of $P_y$ is almost uniform along the *x*-direction (see Sec. II, Fig. S1 of the Supplemental Material [38]). When one kind of charge carrier dominates the doping (here holes at $V_g = -5\ V$), $P_{yL}$ is principally counterbalanced by $P_{yH}$; while at the CNP ($V_g = 0\ V$), only $P_{yD}$ counterbalances $P_{yL}$. We observe that the total charge carrier flux is rigorously null at the edges while it is not in the central part of the Hall bar. Figures 3(c)–3(f) show the transverse electron and hole electric current densities and charge carrier fluxes for two gate voltages along the same cut, as used previously, for our advanced model and when the electrochemical potentials are identical, i.e., $\gamma_n = \gamma_p = \gamma$. This latter case (indicated by the dotted line) corresponds to the two-channel model. Clearly, the electron and hole electric current densities and charge carrier fluxes cancel out at the edges of the Hall bar with our advanced model but they do not when $\gamma_n = \gamma_p = \gamma$, which is inconsistent for a finite Hall bar.

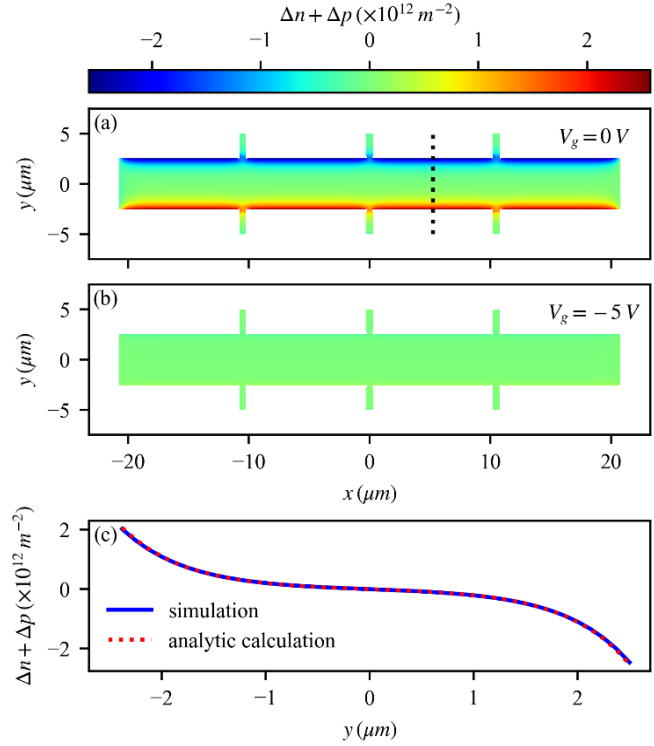

FIG. 4. (a) and (b), color map of $\Delta n + \Delta p$ for two gate voltages. (c) $\Delta n + \Delta p$ along the transverse cut (black dotted lines) represented on (a). The blue curve is the analytical expression (Eq. 11) and the red dotted lines is the simulated curve.

Figures 4(a) and 4(b) show the variation of doping $\Delta n + \Delta p = n - n_{eq} + p - p_{eq}$ for two gate voltages to focus on the occurrence of accumulation and depletion areas near the edges of the Hall bar. These areas are the sources of the transverse electron and hole diffusion fluxes $P_{yD}$. At low bias in the ambipolar regime ($V_g = 0\ V$), these areas are perfectly opposite in amplitude with respect to the plane $y = 0$ [Fig. 4(a)]. At $V_g = -5\ V$, where hole doping is dominant, these areas are negligible. It can be demonstrated that the shape of $\Delta n + \Delta p$ along the transverse direction fits very well at a low bias current and a low magnetic field ($\mu B \ll 1$) with the following expression given (see Sec. II of the Supplemental Material [38]) [blue curve in Fig. 4(c)]

$$\Delta n + \Delta p = \mu^2 B \frac{\partial V_{gr}}{\partial x} \left( n_{eq} \frac{L_n}{D_n} \frac{sh\left(\frac{y}{L_n}\right)}{ch\left(\frac{W}{2L_n}\right)} + p_{eq} \frac{L_p}{D_p} \frac{sh\left(\frac{y}{L_p}\right)}{ch\left(\frac{W}{2L_p}\right)} \right) \quad (11)$$

With $D_{n(p)}$ as the diffusion coefficients of electrons (holes) which is written as

$$D_{n(p)} = \eta \frac{\pi \hbar^2 v_f^2 \mu}{e \int_{E_{CNP}}^{-\eta\infty} 2|E - E_{CNP}||\frac{\partial f}{\partial E}| dE} n_{eq}(p_{eq}) \quad (12)$$

And $L_{n(p)} = \sqrt{D_{n(p)} \tau_r}$ is the diffusion length of the charge carriers (see Supplementary material [38] Sect. II). Equation 11 shows that the size of the accumulation and depletion areas are given by the diffusion length and their amplitude increases with magnetic field, bias current and charge carrier mobility.

**B. Bias current effect**

In this section, we focus on the effect of the bias current on the magnetic field sensitivity shape. We performed simulations on Hall bars having a $\mu = 2 \, m^2/(V.s)$ and two different widths, $W = 1 \, \mu m$ and $W = 5 \, \mu m$. We used $W' = 2W$ and $W_h = W/10$ to ease the comparison with the two-channel model, which was established for an electrodeless Hall bar. Figure 5 shows simulated $S_I(V_g)$ and $R_L(V_g)$ for two different bias current values, 1 µA (black lines) and 200 µA (red lines), and the two widths. First, we observe that the position of the CNP, where $R_L$ is maximum and $S_I$ cancels out, shifts toward positive voltage values as the bias current increases. For the largest Hall bar, the maximum of $R_L$ slightly decreases and the overall shape of $S_I$ is not affected. For the smallest Hall bar, the maximum resistance strongly decreases and the shape of $S_I$ is modified with the current. In particular, the amplitudes of the sensitivity extrema decrease and their gate voltage separation increases. These observations are very similar to our experimental ones (Fig. 1) and those reported elsewhere [6,11,17–19].

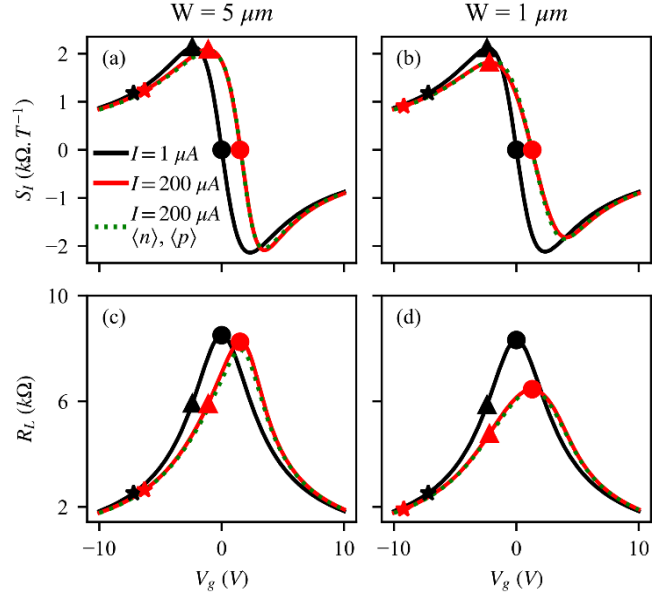

FIG. 5. (a), (b) Simulated $S_I(V_g)$ for 5 µm and 1 µm width Hall bars at 2 different bias current values, 1 µA (black curve) and 200 µA (red curve). (c), (d) Corresponding simulated $R_L(V_g)$. The green dotted lines correspond to curves calculated using equation (1) for $S_I$ and $R = L/We\mu(\langle n \rangle + \langle p \rangle)$ for $R_L$ with $\langle n \rangle$ and $\langle p \rangle$ the simulated average values of *n* and *p* in the Hall bar.

To understand the CNP shift in Fig. 6, the electrostatic potential $V_{gr}$, $n$ and $p$ are represented along the *x*-axis at $y = 0 \, \mu m$, for the two considered bias currents and three gate voltages corresponding to strong positive doping (black and red star symbols), doping where $S_I$ is maximum (black and red triangular symbols), and doping where $S_I = 0$ (black and red circular symbols), i.e., at the CNP (only the 5-µm-width Hall bar is considered, but results are similar for the 1-µm-width Hall bar; see Sec. III, Fig. S3 of the Supplemental Material [38]). We observe that $V_{gr}$ decreases almost linearly along the bar, whatever the bias current value. Regarding the doping, at low bias current, it is uniform in the channel and, as $V_{gr}$ is of the order of a few tenths of mV, it depends only on the gate voltage, $Q_{gr} \approx -C_g V_g$ [see Eq. (4) and Figs. 6(a)–6(c)]. Hence, the CNP gate voltage position corresponds to $Q_{gr} =$ 0 is $V_{CNP} \approx 0\,V$, where $n = p$ and they are minimum in all the channels [Fig. 6(c)]. At a low bias current, our approach and the two-channel model are equivalent, confirming that the two-channel model is a good approximation in this case. Concerning a high bias current, $V_{gr}$ is of the order of magnitude of $V_g$. Hence, the doping is no longer uniform [Figs. 6(d)–6(f)]. For a negative value of $V_g$, the electric charge and, hence, the doping increase near the drain electrode where the gate voltage is added to the drain voltage, $Q_{gr} = -C_g(V_g - V_{gr})$ [see Eq. (4) and Figs. 6(d) and 6(e)]. For a positive value of $V_g$, the drain voltage and gate voltage have an opposite role, i.e., a p–n junction appears in the channel [Fig. (6f)] whose spatial position can be evaluated depending on $V_g$ and $V_{gr}$ using Eq. (4) with $Q_{gr} = 0$. Thus, for a given $V_g$, the x-position of the p–n junction is obtained when $V_{gr}(x) = V_g$. Therefore, observing that the CNP corresponds to the formation of the p–n junction in the middle of the channel where $V_{gr}(x) \approx V_D/2$ [Fig. 6(f)], we deduce that $V_{CNP} \approx V_D/2 = R_{max}I/2$ with $R_{max}$ as the two probes maximum resistance. This conclusion agrees with previous experimental works [18]. It clearly confirms that the CNP gate voltage position is directly related to the bias current due to a doping modulation inside the device, as reported for graphene field effect transistors [23,24].

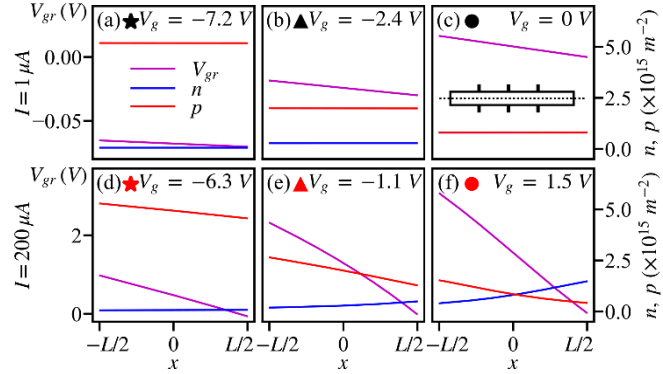

FIG. 6. Charge carrier doping $n$ (blue curves) and $p$ (red curves), electrostatic potential $V_{gr}$ (violet curves) along $x$ for $y = 0\,\mu m$ for the $5\,\mu m$ width Hall bar at three different gate voltage values (see star, triangular and circular symbols on Fig. 3) and for two bias current values, $1\,\mu A$ (a), (b), (c) and $200\,\mu A$ (d), (e), (f). Insert Fig. 6(c): location of the profiles on the Hall bar.

Regarding the bias current-induced evolution of the shape of $S_I$ and $R_L$, in particular, for the smallest Hall bar [Figs. 5(b) and 5(d)], it will be demonstrated in the following that it is a direct consequence of the occurrence of the accumulation and depletion areas near the edges of the Hall bar, as discussed in Sec. A. Their size is of the order of the charge carrier diffusion length $L_{n(p)}$, which is identical for electrons and holes at the CNP and equals 616 nm with the parameters used for the simulation. To obtain insights into this phenomenon, Figs. 7(a)–7(h) show color maps of $n + p$ and $\Delta n + \Delta p$ at gate voltages maximising $S_I$ in the hole regime (black and red triangular symbols in Fig. 5) for both widths and both bias current values, while Figs. 7(i)–7(l) display profiles along the y-axis of $n + p$ and $\Delta n + \Delta p$ in the central part of the Hall bar (at $x = 0\,\mu m$). At a low bias current and for both Hall bar widths, the hole accumulation and depletion areas are uniform along the Hall bar and opposite in amplitude with respect to the plane $y = 0$ (Fig. 7(b) and 7(f)) which is consistent with the observation that the electric charge, which is principally controlled by the gate voltage at low bias current ($Q_{gr} \approx -C_g V_g$, see Eq. 4), must remain uniform. However, their amplitudes are negligible compared with the average value of the charge carrier doping [Figs. 7(a), 7(b), 7(i), 7(e), 7(f), and 7(k)], which is thus almost uniform inside both Hall bars, as we can observe in Figs. 7(a) and 7(e). In Figs. 7(i) and 7(k), a focus is made on the central part of the Hall bar (at $x = 0$ µm); we clearly observe that the average total charge carrier doping $\langle n + p \rangle$ along the y-direction [marked by a black transverse dotted line in Figs. 7(a) and 7(e)] is equal in both Hall bars at a low bias current [red dotted lines in Figs. 7(i) and 7(k)]. It takes the value $2.28 \times 10^{15}\,m^{-2}$, which is uniform in the Hall bar [Figs. 7(a) and 7(e)]. At a high bias current, the situation is more complex. For the largest Hall bar, we observe that the accumulation and depletion areas are almost opposite in amplitude with respect to the plane where $y = 0$ but they are not uniform. Their amplitudes increase toward the source electrode [Fig. 7(d)] and are also more pronounced than at a low bias current, as expected [Figs. 7(c), 7(d), and 7(j)]. However, the accumulation and depletion areas do not strongly affect the total charge carrier doping inside the Hall bar [Fig. 7(c)], whose nonuniformity is mainly induced by the amplitude of $V_{gr}$, as explained in the previous paragraph [$Q_{gr} = -C_g(V_g - V_{gr})$, see Eq. (4)]. Figure 7(j) provides a focus on the central part [Fig.

7(c)]. As explained for the low bias current case, the almost opposite amplitudes of the accumulation and depletion areas still lead to an average $\langle n + p \rangle$ along the $y$-direction that almost equals ($2.32 \times 10^{15}\ m^{-2}$) the one at the low bias current [red dotted lines in Figs. 7(i) and 7(j)]. Therefore, since $S_I$ is inversely proportional to $n + p$ in the central part of the Hall bar, its amplitude barely varies at a high bias current for the largest Hall bar [Fig. 5(a)]. For the smallest Hall bar, Fig. 7(h) shows that the accumulation and depletion areas are nonuniform and have different amplitudes with a strongly dominant accumulation area on the bottom edge of the Hall bar, whose amplitude is now of an order of magnitude of the total average charge carrier doping [Fig. 7(h)]. Consequently, an important modification of the shape of $n + p$ along the Hall bar compared with the case at low bias current is observed [Figs. 7(e) and 7(g)]. This phenomenon relates to the amplitude of $L_{n(p)}$, which is of an order of magnitude of the Hall bar width leading to an important spreading of the accumulation and depletion areas inside the Hall bar. This results in a strong increase in the total charge carrier doping inside the Hall bar [Fig. 7(g)] and, as we observe in Fig. 7(l), $\langle n + p \rangle$ averaged along the $y$-direction [red line in Fig. 7(l)] increases strongly to $2.8 \times 10^{15}\ m^{-2}$. Hence, a large decrease of $S_I$ is observed [Fig. 5(b)]. It is possible to quantitatively reproduce the shape of $S_I$ as a function of gate voltage and bias current using the average simulated doping values $\langle n \rangle$ and $\langle p \rangle$ on the entire Hall bar combined with Eq. (1) [green dotted lines in Figs. 5(a) and 5(b)]. Thus, this analysis demonstrates the key role of the spatial profiles of the charge carrier doping $n$ and $p$ inside the Hall bar, which depends on bias current, gate voltage, and Hall bar geometry.

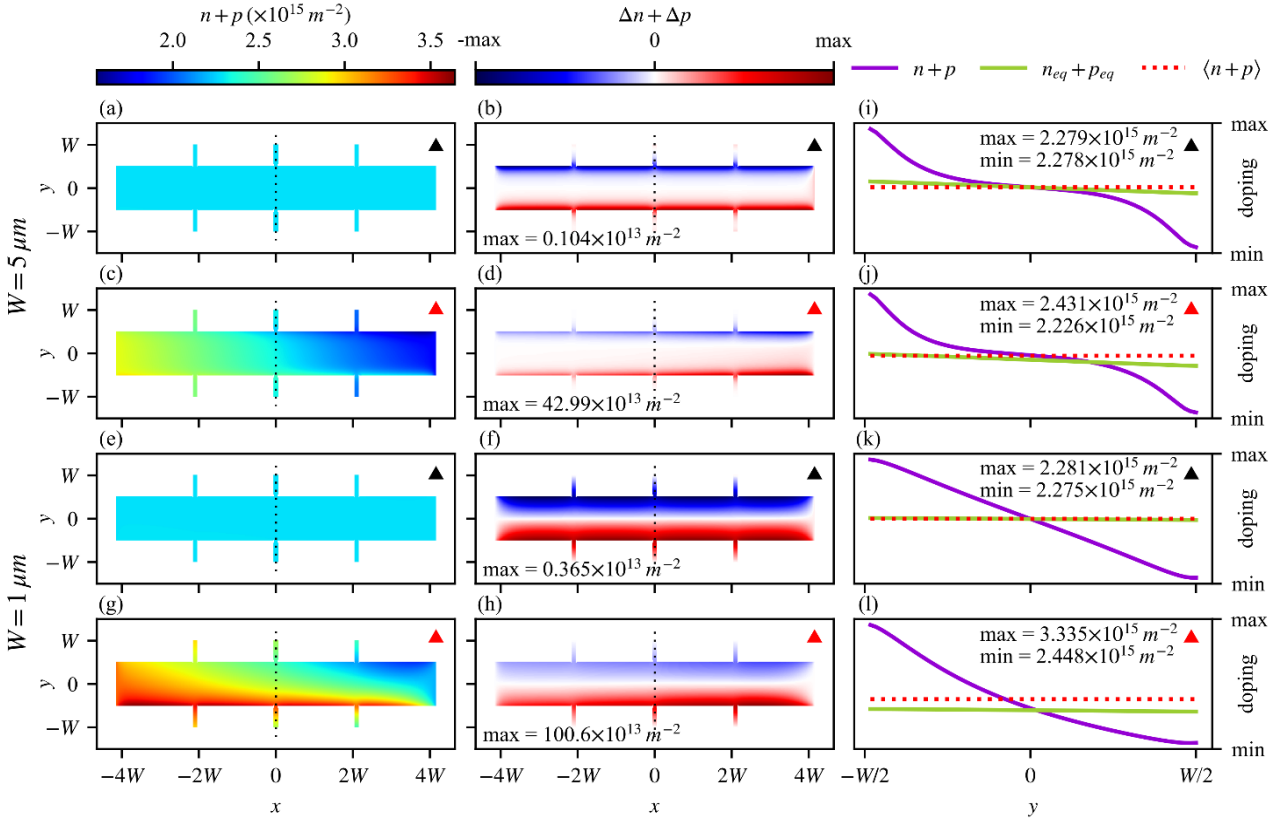

FIG. 7. Colour maps of $n + p$ and $\Delta n + \Delta p$ at the maximum amplitude of $S_I$ in the hole regime for the large Hall bar at low bias current (a) and (b) and high bias current (c) and (d) and for the small Hall bar at low bias current (e) and (f) and high bias current (g) and (h). Transverse cut of $n + p$ in the central part of the Hall bar (black dotted lines on figures (a)-(h)) for the large Hall bar at low bias current (i) and high bias current (j), and for the small Hall bar at low bias current (k) and high bias current (l).

Concerning $R_L$, a similar analysis can be undertaken with maps of $n + p$ at $V_g = V_{CNP}$. Figures 8(b) and 8(f) show that at a low bias current and for both Hall bar widths, the accumulation and depletion areas are perfectly opposite in amplitude with respect to the plane where $y = 0$ As for $S_I$, it leads to a uniform and minimum average doping along the two Hall bars equal to $1.62 \times 10^{15}\ m^{-2}$ [red dotted lines in Figs. 8(i) and 8(k)]. For the largest Hall bar at a high bias current, as explained previously, we still observe opposite accumulation and depletion areas [Fig. 8(d)] with minimum doping in the center of the Hall bar corresponding to the apparition of the p–n junction [Fig. 8(c)]. In addition, $\langle n + p \rangle$ along the $y$-direction in the central part is equal to $1.65 \times 10^{15}\ m^{-2}$ [black dotted lines in Fig. 8(c)], a value barely equal to the one at a low bias current [Figs. 8(i) and 8(j)]. As a result, the resistance, which is inversely proportional to $n + p$, is nearly equal for both low and high bias currents. For the smallest Hall bar, the shape of $n + p$ is strongly affected by the bias current with a dominant accumulation area at the bottom edge of the Hall bar, resulting in an increase in the average value of $n + p$ compared with the value at a low bias current [red lines in Figs. 8(k) and 8(l)], which is consistent with the maximum resistance decreases at a high bias current [Fig. 5(d)]. In the same way as for $S_I$, it is possible to quantitatively reproduce the shape of $R_L$ as a function of gate voltage and bias current using the average simulated doping values $\langle n \rangle$ and $\langle p \rangle$ on the entire Hall bar and combined with the equation $R = L/We\mu(\langle n \rangle + \langle p \rangle)$ (green dotted lines in Fig. 5).

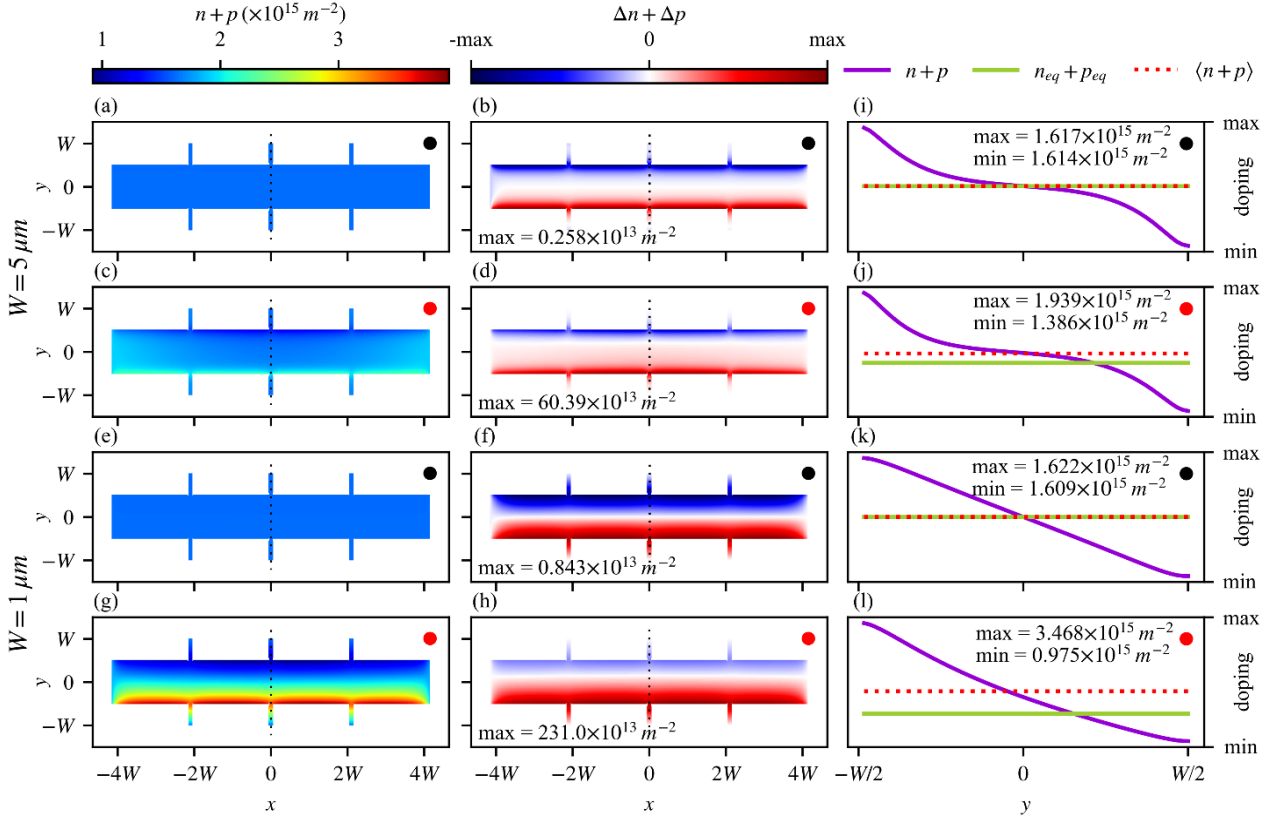

FIG. 8. Colour maps of $n + p$ and $\Delta n + \Delta p$ at the maximum amplitude of $R_L$ for the large Hall bar at low bias current (a) and (b) and high bias current (c) and (d) and for the small Hall bar at low bias current (e) and (f) and high bias current (g) and (h). Transverse cut of $n + p$ in the central part of the Hall bar (black dotted lines on figures (a)-(h)) for the large Hall bar at low bias current (i) and high bias current (j), and for the small Hall bar at low bias current (k) and high bias current (l).

To conclude this section, we performed simulations with an increased recombination-generation rate $R'$ (using $k' = 10k$) in order to clearly demonstrate that when the accumulation and depletion area amplitudes are decreased, $S_I$ and $R_L$ gate voltage and bias current dependences are less affected. We focused on the smallest GHSs where the effect is more pronounced. Figure 9(a) shows that, with $k' = 10k$, the shape of $S_I$ is less affected at a high bias current (blue curve), its maximum amplitude does not decrease strongly compared with the case at a low bias current. Figure 9(b) shows the profile of $n + p$ along a transverse cut at $x = 0 \, \mu m$ for $k$ and $k'$. It is clear that the accumulation and depletion areas are less pronounced when the recombination rate is high (for $k'$, blue curve), resulting in an average value of $n + p$ closer to the equilibrium value. This observation is consistent with the decrease of $L_{n(p)}$, which is equal to 195 nm for $k'$. Finally, this comparison of our model and the two-channel model clearly points out the shortcomings of the two-channel model. It reveals the important role of the accumulation and depletion areas on the electrical characteristics of GHSs.

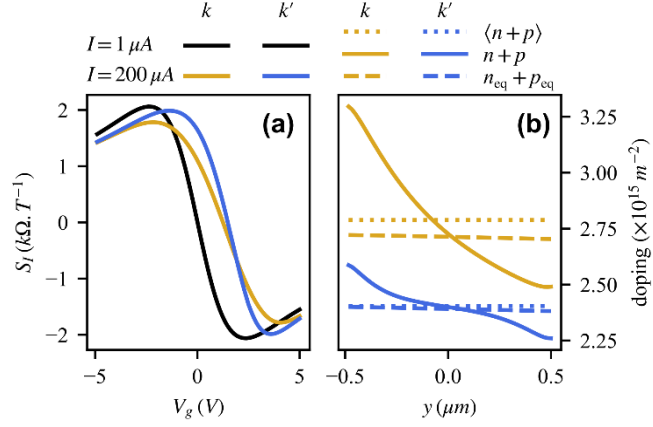

FIG. 9. (a) Magnetic field sensitivity $S_I$ for two values of the bias current and two values of $k$. (b) Profile of $n + p$ along a transverse cut at $x = 0\ \mu m$ at high bias current for two values of $k$.

## C. Electron-hole puddle effect

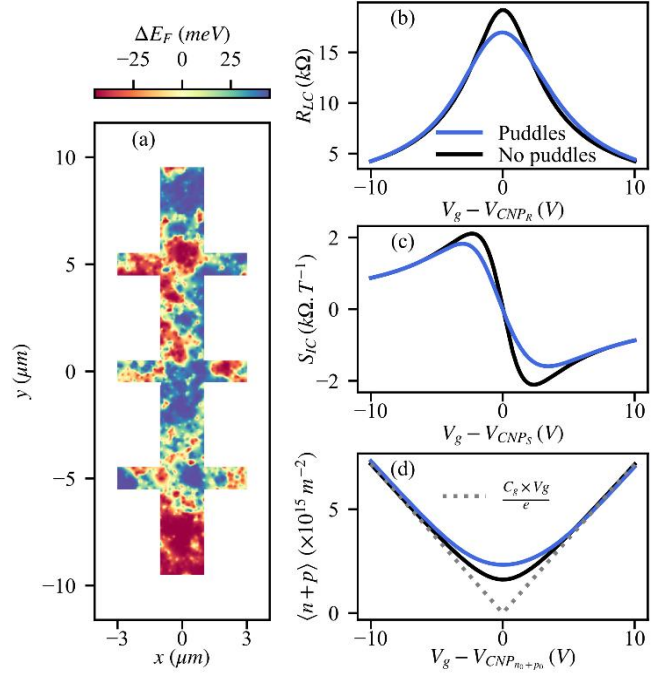

FIG. 10. (a) Colour map of the Fermi level fluctuations in the Hall bar induced by the random gate voltage fluctuations. The parameters used to generate the map were $\Delta E_{fmax} = 45\ meV$, $20\ nm < r_{0i} < 50\ nm$ and $n_{0l} = 3 \times 10^9\ cm^{-2}$. (b) and (c) Simulated longitudinal resistance $R_{LC}$ and magnetic field sensitivity $S_{IC}$ with puddles (blue curves) and without puddles (black curves). (d) Total average doping $\langle n + p \rangle$ as a function of $V_g$ with puddles (blue curves) and without puddles (black curves). The GHS being now strongly inhomogeneous, the positions of the CNP are slightly different for $R_{LC}$, $S_{IC}$ and $\langle n + p \rangle$ and are noted $V_{CNP_R}$, $V_{CNP_S}$ and $V_{CNP_{n_0+p_0}}$.

Figure 10(a) shows a typical map of the Fermi-level fluctuations generated using Eq. (8) at $V_g = 0\ V$. The parameters used are $\Delta E_{frmax} = 45\ meV$, $r_{0i}$ ranging from 20 nm to 50 nm and $n_{0l} = 3 \times 10^9\ cm^{-2}$. The Fermi-level fluctuations have a root mean square of 31 meV and induce a mean residual charge carrier doping $\langle n + p \rangle$ of $2.32 \times 10^{11}\ cm^{-2}$, a value above the thermal one equal to $1.61 \times 10^{11}\ cm^{-2}$. The FWHM of the autocorrelation map function allows for an extraction of a mean puddle size of 1 μm, meaning the puddles span the Hall bar. Figures 10(b) and 10(c) show the compensated longitudinal resistance $R_{LC}$ and magnetic field sensitivity $S_{IC}$ simulated with and without the puddles for $W = 2\ \mu m$, $W' = 3W$, $W_h = W/2$ and $\mu = 1\ m^2/(V.s)$. While the overall shape of $R_{LC}$ and $S_{IC}$ is preserved, the main effect of the introduction of the puddle is to decrease both the resistance and sensitivity near the CNP, as expected. For large gate voltages, corresponding to high doping and single charge carrier type behavior (at $|V_g| \gtrsim 5\,V$), there are no significant changes as the Fermi-level fluctuations become negligible [26] [Fig. 10(d)]. The chosen values of $n_{0l}$ and $r_{0i}$ were based on experimental observations of the puddles [31,32,35] and the requirement that the Boltzmann formalism must remain valid. This implies that the conductivity should not vary on a length scale greater than the mean free path $l_m$ [36], which mathematically means that $\sigma/\|\vec{\nabla}\sigma\| > 5 l_m$ with $l_m = \hbar\sqrt{\pi}\mu(n+p)/e(\sqrt{n}+\sqrt{p})$. Under these conditions, 95% of the Hall bar surface satisfies the criterion.

## V. ANALYSIS OF THE EXPERIMENTAL RESULTS

For a proper comparison between the experiments and simulations, the dimensions of the measured GHSs were obtained with a Nano-Observer atomic force microscope from CSI (see Fig. S4 of the Supplemental Material [38]). In addition, it is important to note that the input parameters $\mu$, $v_f$, $\Delta E_{frmax}$ and $n_{0l}$, which correspond to the charge carrier mobility, the Fermi velocity, the maximum amplitude of the fluctuations of the Fermi level, and the Lorentzian density, respectively, were evaluated at a small bias current and the base temperature of the experiment (i.e., 200 K for CVD-GHSs and 300 K for HOPG and hBN-GHSs). Then, these values were used to model the behavior of the GHSs at a high bias current while keeping the temperature of the devices at the base temperature of the experiment, meaning that the Joule effect heating was neglected. As will be detailed in the following, this procedure allows us to reproduce quantitatively the experimental data, which is a strong argument to neglect the Joule effect at a high bias current. However, to justify this hypothesis beyond any doubt, numerical simulations of the temperature increase induced by the Joule effect heating were performed and confirmed that this effect can be neglected. The detailed procedure is explained in Part V of the Supplemental Material [38].

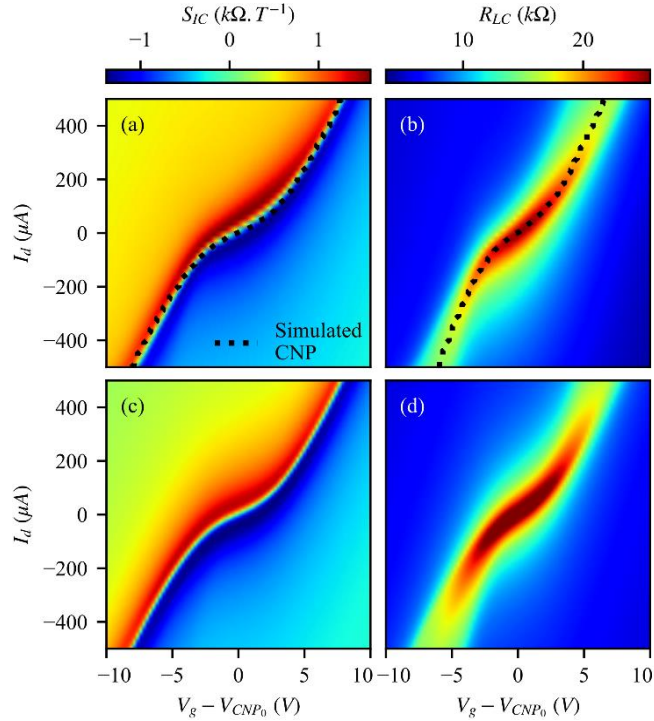

FIG. 11. Experimental maps of (a) $R_{LC}$ and (b) $S_{IC}$ for CVD-GHS ($B = 1\,T$, $T = 200\,K$), compared with simulations of (c) $R_{LC}$ and (d) $S_{IC}$ (see Table 1). The black dot lines in figures (a), (b) represent the simulated $V_{CNP}$ extracted from (c), (d). $V_{CNP0}$ is the CNP gate voltage position at the smallest bias current.

## A. Analysis of the CVD GHS

Figure 11 shows the experimental and simulated maps of $R_{LC}$ and $S_{IC}$ of the CVD-GHSs as a function of bias current and gate voltage at $T = 200$ K and $B = 1$ T. Table I lists the parameters used for the simulation. For this device, in addition to long-range scatterers, short-range scatterers were introduced to reproduce the sublinear behavior of the longitudinal conductance as a function of the gate voltage [see Fig. S6(a) of the Supplemental Material [38]]. Short-range scatterers induce resistivity $\rho_{sr}$ independent of the electron and hole doping [39]; hence, the short-range scatterer charge carrier mobility is written as $\mu_{sr}(E) = \pi(\hbar v_f)^2/e\rho_{sr}|E - E_{CNP}|^2$. The simulated data effectively agree with the experimental ones: the amplitudes of $R_{LC}$ and $S_{IC}$ are almost perfectly reproduced together with their overall shapes, especially the CNP shift (see Fig. S6 of the Supplemental Material [38]). Indeed, the dotted lines that appear in Figs. 11(a) and 11(b) represent the $V_{CNP}$ extracted from the simulations. The mean residual charge carrier doping $\langle n + p \rangle$ is about $2.3 \times 10^{15}\ m^{-2}$, a value much larger than the thermal one (i.e., $7.4 \times 10^{14}\ m^{-2}$), which means that there is an extensive number of electron–hole puddles. This observation is coherent with the presence of impurities and defects observed from the Raman measurements [Fig. 1(a)] and the low mobility and large recombination-generation rate used for the simulations (Table I). For this device, accumulation and depletion areas are negligible due to a diffusion length of 39 nm. The very good agreements between the simulations and experiments show that our method of introducing electron–hole puddles is sufficiently effective for analyzing large samples with an extensive number of impurities.

TABLE I. Parameters used for the comparison between experiments and simulations.

| Parameters | CVD | HOPG | hBN |
|---|---|---|---|
| $t_{ox}(nm)$ | 105 | 280 | 300 |
| $v_f(10^6\ m/s)$ | 1 | 1.2 | 1.4 |
| $\mu_{lr}(m^2/Vs)$ | 0.7 | 1.5 | 5 |
| $\rho_{sr}(\Omega)$ | 1000 | 0 | 0 |
| $n_{0l}(10^9 cm^{-2})$ | 3 | 3 | 3 |
| $r_{0i}(nm)$ | 50 − 80 | 60 − 100 | 60 − 100 |
| $\Delta E_{fmax}(meV)$ | 85 | 18 | 15 |
| $k(m^2/s)$ | $10^{-2}$ | $3.10^{-4}$ | $1.10^{-5}$ |

## B. Analysis of the HOPG and hBN GHS

Figure 12 shows the experimental and simulated $R_{LC}$ and $S_{IC}$ of the HOPG and hBN-GHSs for two bias current values, 1 µA (black lines) and 100 µA (red lines) for HOPG-GHSs and 2 µA (black lines) and 200 µA (red lines) for hBN-GHSs at $T = 300\ K$ and $B = 0.1\ T$. With the exception of the $R_{LC}$ amplitude of hBN-GHSs at a high bias current, the simulated data effectively agrees with the experimental ones. Importantly, the maxima of $S_{IC}$ reaches $3\ k\Omega/T$ and $4\ k\Omega/T$ at a low bias current for HOPG and hBN-GHSs, respectively. These values are larger than $2\ k\Omega/T$, the value expected at 300 K for a pristine graphene considering $v_f = 10^6\ m/s$ (the most frequently used value of the Fermi velocity). This discrepancy means that the charge carrier density must be lower, i.e., the density of states must be lower and the Fermi velocity higher. Several studies of the electronic properties of graphene using angle-resolved photoemission spectroscopy (ARPES), STM, or Terahertz measurements have shown that $v_f$ can reach values as high as $1.5 \times 10^6\ m/s$ [40–42]. Such renormalization of the Fermi velocity is mainly due to poorly screened electron-electron interactions in suspended graphene [43] and h-BN encapsulated graphene [40]. Accordingly, our simulations lead to $v_f = 1.2 \times 10^6\ m/s$ and $v_f = 1.4 \times 10^6\ m/s$ for HOPG and hBN-GHSs, respectively. These values agree with previous works [40–42]. However, despite the fact that the total charge carrier density must be lower to explain the $S_{IC}$ amplitude, it is still necessary to introduce electron–hole puddles to reproduce the measured $R_{LC}$ and $S_{IC}$ gate voltage dependences, especially at a high bias current. For HOPG-GHSs, the combination of a higher Fermi velocity and electron–hole puddles leads to a mean residual charge carrier doping $\langle n + p \rangle$ of $1.2 \times 10^{15}\ m^{-2}$, a value slightly superior to $1.1 \times 10^{15}\ m^{-2}$, the value expected without puddles. For hBN-GHSs, the mean residual charge carrier doping of $8.3 \times 10^{14}\ m^{-2}$ remains almost unaffected, although the presence of electron–hole puddles plays a role in the amplitude of the resistance at a high bias current. Both values are lower than $1.61 \times 10^{15} \, m^{-2}$, the expected value for pristine graphene with $v_f = 10^6 \, m/s$. It is noteworthy that it was not necessary to introduce a Fermi velocity renormalization for the CVD-GHSs because the number of charge carriers is larger, meaning that the electron-electron interaction is efficiently screened. Concerning the shape of $S_{IC}$, its large modulation at a high bias current for both devices is effectively reproduced by our model. This means that the accumulation and depletion areas play an important role in these devices, as described in Sec. B. Indeed, unlike CVD-GHSs, the diffusion lengths at the CNP are 345 nm and 4.32 µm for HOPG and hBN-GHSs, respectively, which is of the order or larger than the device width, as a direct consequence of larger mobility and a lower recombination rate for both samples compared with the CVD-GHSs (Table I).

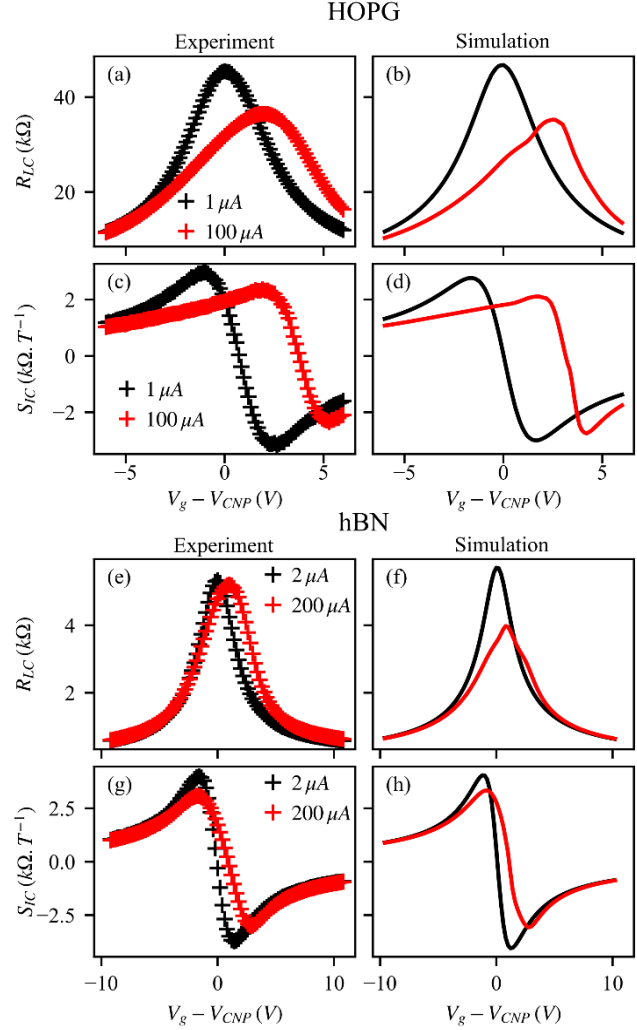

FIG. 12. Experimental (a) and (b) and simulated (e) and (f) $R_{LC}$ and $S_{IC}$ of the HOPG-GHS (black lines 1 µA, red lines 100 µA). Experimental (c) and (d) and simulated (g) and (h) $R_{LC}$ and $S_{IC}$ of the hBN-GHS (black lines 2 µA, red lines 200 µA). Both samples were measured at $T = 300 \, K$ and $B = 0.1 \, T$.

Thus, the good agreement between our advanced model and experiments highlights the importance of considering distinct electron and hole Fermi levels and recombinationgeneration processes in order to address the evolution of $S_{IC}$ at a high bias current. In addition, a coherent evolution of the mean residual charge carrier, mobility, and recombination rate is observed for the three samples (Table I). Indeed, as expected, the mobility increases as the mean residual charge carrier decreases [26,27], and the recombination rate decreases as the Fermi velocity and the quality of the device increase [29,30]. Moreover, the thermal charge carrier statistics used in our model allow us to emphasize the importance of considering the renormalization of the Fermi velocity to explain the amplitude of $S_{IC}$. Regarding the amplitude of $R_{IC}$, which our model does not reproduce quantitatively for hBN-GHSs, we believe that it may be related to the size of the electron–hole puddles, whose amplitude is similar to the width of the device, and their shapes, which have been shown to elongate at the edges, a feature that our method cannot take into account quantitatively [35]. This may also be linked to the use of linear recombination-generation rates, whereas nonlinear processes are expected when the electron and hole populations are far from equilibrium near the edges of the Hall bar, particularly for GHSs with high charge carrier mobility such as hBN-GHSs.

## VI. CONCLUSION

In summary, we have developed a comprehensive numerical model with several improvements over the two-channel model: (i) our model considers the effect of temperature and different processes of charge carrier scattering, (ii) our model takes into account the spatial modulation of the charge carrier doping as a function of bias current, gate voltage, and geometry by using distinct Fermi levels for electrons and holes and a local field effect model for electrostatic doping, (iii) our model accounts for charge carrier inhomogeneities introduced by substrates and contamination by using a semiempirical method to locally introduce electron–hole puddles, unlike previous works where the influence of the puddles is treated on average [22,37]. Consequently, our advanced model can quantitatively reproduce the galvanomagnetic properties of GHSs at different qualities and different conditions of biasing, contrary to the two-channel model. In addition, an in-depth understanding of the operation principles and limitations of GHSs is obtained. In particular, our model reveals how accumulation and depletion areas that form near the edges of the Hall bar in the ambipolar regime can affect and degrade the performance of GHSs with widths of the order of the charge carrier diffusion length (see Fig. S7 of the Supplemental Material [38]). It would, therefore, be interesting to compare the predicted performance of GHSs that are a few hundred nanometers wide with their actual performance. Additionally, we demonstrated how the substrate and, more generally, the electrostatic environment of the graphene can affect the GHSs performance through the variation of the Fermi velocity. This means that the use of substrates or encapsulation materials with low relative dielectric constants [42] or the use of suspended graphene [43] should improve performance. It is worth noting that Fermi velocity can also be modulated by in-plane uniform strain [44]. It could constitute another method to increase the magnetic field sensitivity of GHSs.

Despite the good agreements between the simulations and the experiments presented in the paper, further improvements can still be made. Indeed, our model has three input parameters: the value of the Fermi velocity, the expression of the recombination rate, and the map of the electron–hole puddles. First, regarding the value of the Femi velocity, it will be very helpful to perform ARPES measurements on our samples to obtain a precise estimation and confirm its role. Second, the linear expression of the recombination-generation rate used in our model may not be relevant when the electron and hole populations are far from the equilibrium, especially for small high-quality samples biased with high bias current. Better agreements between simulations and experiments, particularly concerning the amplitude of the resistance, should be obtained for high-quality samples if realistic and nonlinear processes are included, such as the recombination-generation induced by optical phonons [30]. Future work will be carried out in this sense. Regarding the map of the electron–hole puddles, it should be helpful to integrate real maps measured under various conditions and different substrates using, for example, nearfield photocurrent nanoscopy [35], especially to obtain proper modeling of the electron–hole puddle size and shape inside the Hall bar [35]. This last point would allow us to obtain improved agreements between simulations and experiments, particularly for small high-quality samples where the exact map of electron–holes puddles, whose sizes are of the order of magnitude of the width of the Hall bar, can have a strong influence on the shape of $R_{LC}$ and $S_{IC}$. Finally, it is noteworthy that, recently, a Hall sensor made with a heterostructure of h-BN/MoS$_2$ was realized [45]. Our advanced model could be very helpful in studying, in detail, the operation of such a sensor since it is straightforward to adapt it to account for a different electronic band structure, charge carrier mobility energy dependence, and recombination generation rate.

## ACKNOWLEDGMENTS

The authors thank Ioannis Paradisanos, Mathieu Pierre, and Walter Escoffier for fruitful discussions. This study has been supported through the EUR grant NanoX n° ANR-17-EURE-0009 in the framework of the "Programme des Investissements d'Avenir".

# Supplemental material

# Performance of graphene Hall effect sensors: role of bias current, disorder and Fermi velocity

Lionel Petit[1], Tom Fournier[1,2], Géraldine Ballon[1], Cédric Robert[1], Delphine Lagarde[1], Pascal Puech[2], Thomas Blon[1], and Benjamin Lassagne[1]

[1] Université de Toulouse, INSA-CNRS-UPS, LPCNO, 135 Av. Rangueil, 31077 Toulouse, France
[2] Centre d'Elaboration des Matériaux et d'Etudes Structurales (CEMES), UPR8011 CNRS, Université Toulouse 3, 31055 Toulouse, France
E-mail: lassagne@insa-toulouse.fr

### I. Random gate voltage fluctuations

To convert the Fermi level fluctuations into gate voltage fluctuations we used Eq. 4 at zero gate voltage and zero bias current. Hence, the electrochemical potentials of electrons and holes are null, $\gamma_n = \gamma_p = 0$ and as $W_M = W_{gr}$, $\Delta E_{fr} = eV_{gr}$. Using $Q_0 = 0$, we can write

$$\Delta V_g = \frac{\Delta E_{fr}}{e} + \frac{e}{C_g}\left(n(\Delta E_{fr}) - p(\Delta E_{fr})\right) \qquad (s1)$$

### II. Comparison with the two-channel model

At low magnetic field, low bias current and for a constant $\mu$, the conductivity components write

$$\sigma_{xx_{n(p)}} \approx \frac{\eta e \mu}{\pi v_f^2 \hbar^2} \int_{E_{CNP}}^{-\eta \times \infty} (E - E_{CNP})^2 \frac{\partial f_{n(p)}}{\partial E} dE = en(p)\mu$$

$$\sigma_{xy_{n(p)}} \approx \frac{e\mu^2 B}{\pi v_f^2 \hbar^2} \int_{E_{CNP}}^{-\eta \times \infty} (E - E_{CNP})^2 \frac{\partial f_{n(p)}}{\partial E} dE = \eta en(p)\mu^2 B$$

The electric current densities and the charge carrier flux write in the two-channel model as follow

$$\vec{J}_n = -e\vec{P}_n = -en\mu \begin{vmatrix} \frac{\partial V_{gr}}{\partial x} - \mu B \frac{\partial V_{gr}}{\partial y} \\ \mu B \frac{\partial V_{gr}}{\partial x} + \frac{\partial V_{gr}}{\partial y} \end{vmatrix} \qquad (s2a)$$

$$\vec{J}_p = e\vec{P}_p = -ep\mu \begin{vmatrix} \frac{\partial V_{gr}}{\partial x} + \mu B \frac{\partial V_{gr}}{\partial y} \\ -\mu B \frac{\partial V_{gr}}{\partial x} + \frac{\partial V_{gr}}{\partial y} \end{vmatrix} \quad (s2b)$$

Hence the total electric current density and the total charge carrier flux write

$$\vec{J} = \begin{vmatrix} -e n \mu (n+p) \frac{\partial V_{gr}}{\partial x} + e(n-p)\mu^2 B \frac{\partial V_{gr}}{\partial y} \\ -e(n-p)\mu^2 B \frac{\partial V_{gr}}{\partial x} - e(n+p)\mu \frac{\partial V_{gr}}{\partial y} \end{vmatrix} \quad (s3a)$$

$$\vec{P} = \begin{vmatrix} (n-p)\mu \frac{\partial V_{gr}}{\partial x} - (n+p)\mu^2 B \frac{\partial V_{gr}}{\partial y} \\ (n+p)\mu^2 B \frac{\partial V_{gr}}{\partial x} + (n-p)\mu \frac{\partial V_{gr}}{\partial y} \end{vmatrix} \quad (s3b)$$

In our advanced model, the electric current densities and charge carrier flux write

$$\vec{J}_n = -e\vec{P}_n = n\mu \begin{vmatrix} \frac{\partial \gamma_n}{\partial x} - \mu B \frac{\partial \gamma_n}{\partial y} \\ \mu B \frac{\partial \gamma_n}{\partial x} + \frac{\partial \gamma_n}{\partial y} \end{vmatrix} \quad (s4a)$$

$$\vec{J}_p = e\vec{P}_p = p\mu \begin{vmatrix} \frac{\partial \gamma_p}{\partial x} + \mu B \frac{\partial \gamma_p}{\partial y} \\ -\mu B \frac{\partial \gamma_p}{\partial x} + \frac{\partial \gamma_p}{\partial y} \end{vmatrix} \quad (s4b)$$

Hence, the total electric current density and charge carrier flux expressions are

$$\vec{J} = \begin{vmatrix} \mu \left( n \frac{\partial E_{fn}}{\partial x} + p \frac{\partial E_{fp}}{\partial x} \right) - e(n+p)\mu \frac{\partial V_{gr}}{\partial x} - \mu^2 B \left( n \frac{\partial E_{fn}}{\partial y} - p \frac{\partial E_{fp}}{\partial y} \right) + e(n-p)\mu^2 B \frac{\partial V_{gr}}{\partial y} \\ \mu^2 B \left( n \frac{\partial E_{fn}}{\partial x} - p \frac{\partial E_{fp}}{\partial x} \right) - e(n-p)\mu^2 B \frac{\partial V_{gr}}{\partial x} + \mu \left( n \frac{\partial E_{fn}}{\partial y} + p \frac{\partial E_{fp}}{\partial y} \right) - e(n+p)\mu \frac{\partial V_{gr}}{\partial y} \end{vmatrix} \quad (s5a)$$

$$\vec{P} = \begin{vmatrix} -\frac{\mu}{e} \left( n \frac{\partial E_{fn}}{\partial x} - p \frac{\partial E_{fp}}{\partial x} \right) + (n-p)\mu \frac{\partial V_{gr}}{\partial x} + \frac{\mu^2 B}{e} \left( n \frac{\partial E_{fn}}{\partial y} + p \frac{\partial E_{fp}}{\partial y} \right) - (n+p)\mu^2 B \frac{\partial V_{gr}}{\partial y} \\ -\frac{\mu^2 B}{e} \left( n \frac{\partial E_{fn}}{\partial x} + p \frac{\partial E_{fp}}{\partial x} \right) + (n+p)\mu^2 B \frac{\partial V_{gr}}{\partial x} - \frac{\mu}{e} \left( n \frac{\partial E_{fn}}{\partial y} - p \frac{\partial E_{fp}}{\partial y} \right) + (n-p)\mu \frac{\partial V_{gr}}{\partial y} \end{vmatrix} \quad (s5b)$$

At low bias current, all the first terms of Eqs. s5(a) and s5(b) depending on the gradient of the Fermi levels along $x$ $\left( \frac{\partial E_{fn}}{\partial x}, \frac{\partial E_{fp}}{\partial x} \right)$ can be neglected as they are almost null.

In the following, we consider the device operating at the CNP and supplied with a low bias current, typically few µA. The electron and hole doping spatial variations are expressed as follows

$$n = n_{eq} + \Delta n$$

$$p = p_{eq} + \Delta p$$

$n_{eq}$ and $p_{eq}$ are the electron and hole doping at equilibrium. At low bias current, the doping variations are small, meaning that $\Delta n \ll n_{eq}$ and $\Delta p \ll p_{eq}$.

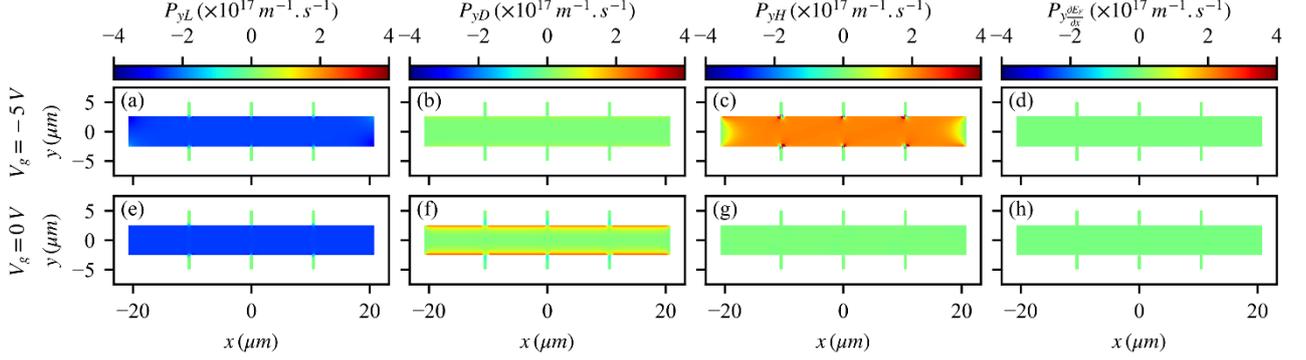

FIG. S1. (a) and (e), Lorentz component of the transverse charge carrier flux, (b) and (f), Diffusion component, (c) and (g) Hall component, (d) and (h) components depending on $\frac{\partial E_{fn}}{\partial x}$ and $\frac{\partial E_{fp}}{\partial x}$ for two gate voltages.

The spatial dependences of the charge carrier flux $\vec{P}_n$ and $\vec{P}_P$ are governed by the following relationships

$$-\vec{\nabla}.\vec{P}_{n(p)} - R = 0 \tag{s6}$$

With $R = k(np - n_{eq}p_{eq})$. At the CNP we assume that $\Delta n = \Delta p$, therefore we can write

$$R \approx k(\Delta n p_{eq} + \Delta p n_{eq}) = \frac{\Delta n}{\tau_r} = \frac{\Delta p}{\tau_r} \tag{s7}$$

With $\tau_r = 1/k(p_{eq} + n_{eq})$ the recombination rate. Injecting the relationships $\vec{P}_{n(p)} = \frac{\eta}{e}\vec{J}_{n(p)}$ and Eq. s7 into Eq. s6, and assuming that $\frac{\partial^2 \gamma_{n(p)}}{\partial x \partial y} = \frac{\partial^2 \gamma_{n(p)}}{\partial y \partial x}$ we can obtain the following relationships

$$\frac{n_{eq}\mu}{e}\left(\frac{\partial^2 \gamma_n}{\partial x^2} + \frac{\partial^2 \gamma_n}{\partial y^2}\right) - \frac{\Delta n}{\tau_r} = 0 \tag{s8a}$$

$$-\frac{p_{eq}\mu}{e}\left(\frac{\partial^2 \gamma_p}{\partial x^2} + \frac{\partial^2 \gamma_p}{\partial y^2}\right) - \frac{\Delta p}{\tau_r} = 0 \tag{s8b}$$

The simulations performed at the CNP ($V_g = 0V$) reveal that in almost the entire Hall bar (Fig. S2):

$$\frac{\partial E_{fn}}{\partial x}, \frac{\partial E_{fp}}{\partial x}, \frac{\partial V_{gr}}{\partial y}, \frac{\partial^2 V_{gr}}{\partial x^2} \approx 0$$

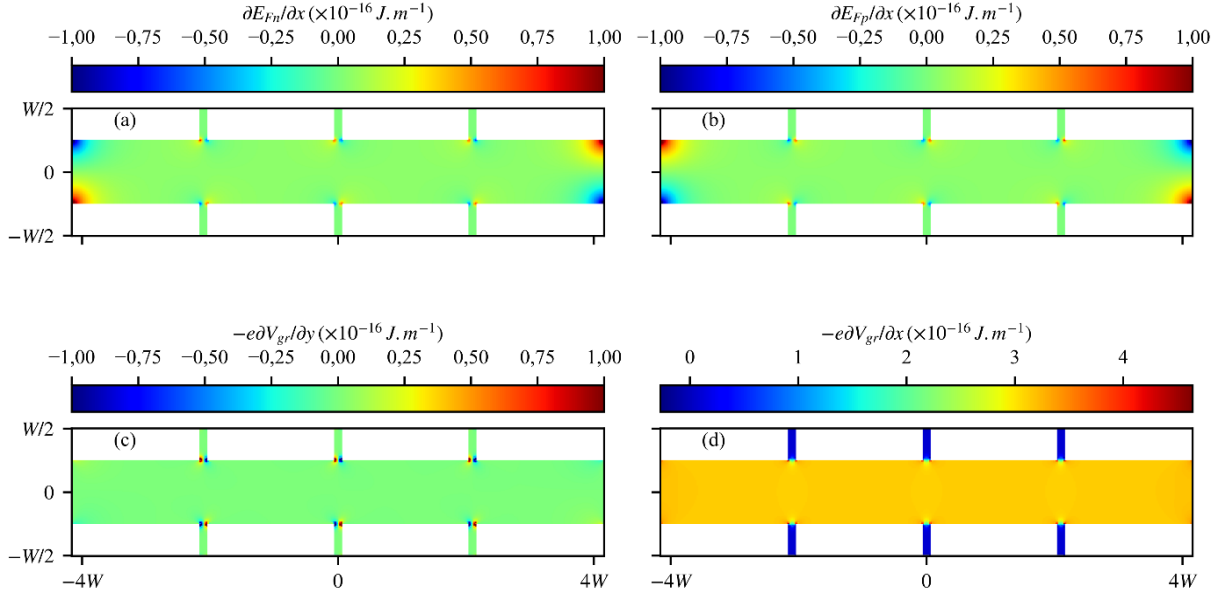

FIG. S2. Color maps of $\frac{\partial E_{fn}}{\partial x}$ (a), $\frac{\partial E_{fp}}{\partial x}$ (b) $\frac{\partial V_{gr}}{\partial y}$ (c) and $\frac{\partial V_{gr}}{\partial x}$ (d).

Hence the Eqs. s8(a) and s8(b) become

$$\frac{n_{eq}\mu}{e}\frac{\partial^2 E_{fn}}{\partial y^2} - \frac{\Delta n}{\tau_r} = 0 \qquad (s9a)$$

$$-\frac{p_{eq}\mu}{e}\frac{\partial^2 E_{fp}}{\partial y^2} - \frac{\Delta p}{\tau_r} = 0 \qquad (s9b)$$

We remind that

$$n = \int_{E_{CNP}}^{\infty} f(E - E_{fn})\frac{2|E - E_{CNP}|}{\pi(\hbar v_f)^2} dE$$

$$p = \int_{-\infty}^{E_{CNP}} \left(1 - f(E - E_{fp})\right)\frac{2|E - E_{CNP}|}{\pi(\hbar v_f)^2} dE$$

Hence, the first derivatives of $n$ and $p$ write

$$\frac{\partial n}{\partial y} = -\frac{\partial E_{fn}}{\partial y} \int_{E_{CNP}}^{\infty} \frac{2|E - E_{CNP}|}{\pi(\hbar v_f)^2} \frac{\partial f}{\partial E} dE \tag{s10a}$$

$$\frac{\partial p}{\partial y} = \frac{\partial E_{fp}}{\partial y} \int_{-\infty}^{E_{CNP}} \frac{2|E - E_{CNP}|}{\pi(\hbar v_f)^2} \frac{\partial f}{\partial E} dE \tag{s10b}$$

And the second derivatives write

$$\frac{\partial^2 n}{\partial y^2} = -\frac{\partial^2 E_{fn}}{\partial y^2} \int_{E_{CNP}}^{\infty} \frac{2|E - E_{CNP}|}{\pi(\hbar v_f)^2} \frac{\partial f}{\partial E} dE + \left(\frac{\partial E_{fn}}{\partial y}\right)^2 \int_{E_{CNP}}^{\infty} \frac{2|E - E_{CNP}|}{\pi(\hbar v_f)^2} \frac{\partial^2 f}{\partial E^2} dE \tag{s11a}$$

$$\frac{\partial^2 p}{\partial y^2} = \frac{\partial^2 E_{fp}}{\partial y^2} \int_{-\infty}^{E_{CNP}} \frac{2|E - E_{CNP}|}{\pi(\hbar v_f)^2} \frac{\partial f}{\partial E} dE - \left(\frac{\partial E_{fp}}{\partial y}\right)^2 \int_{-\infty}^{E_{CNP}} \frac{2|E - E_{CNP}|}{\pi(\hbar v_f)^2} \frac{\partial^2 f}{\partial E^2} dE \tag{s11b}$$

The second terms of the Eqs. s11(a) and s11(b) are negligible, then we can convert the Eqs. s9(a) and s9(b) as follow

$$D_n \frac{\partial^2 n}{\partial y^2} - \frac{\Delta n}{\tau_r} = 0 \tag{s12b}$$

$$D_p \frac{\partial^2 p}{\partial y^2} - \frac{\Delta p}{\tau_r} = 0 \tag{s12b}$$

Where $D_n$ and $D_p$ are the diffusion coefficient of electrons and holes which write

$$D_n = \frac{\pi(\hbar v_f)^2 n_{eq}\mu}{e \int_{E_{CNP}}^{\infty} 2|E - E_{CNP}|\left(-\frac{\partial f}{\partial E}\right) dE} \tag{s13a}$$

$$D_p = \frac{\pi(\hbar v_f)^2 p_{eq}\mu}{e \int_{-\infty}^{E_{CNP}} 2|E - E_{CNP}|\left(-\frac{\partial f}{\partial E}\right) dE} \tag{s13b}$$

We can verify that at $T = 0\ K$, $D_{n(p)} = \hbar v_f \sqrt{\pi n_{eq}(p_{eq})}\mu/2e$. The solutions of Eqs. s12(a) and s12(b) take the following forms

$$\Delta n = Ae^{\frac{y}{L_n}} + A'e^{-\frac{y}{L_n}} \tag{s14a}$$

$$\Delta p = Be^{\frac{y}{L_p}} + B'e^{-\frac{y}{L_p}} \tag{s14b}$$

With $L_{n(p)} = \sqrt{D_{n(p)}\tau_r}$ the diffusion length of electrons(holes). Using the Eqs. s4(a) and s4(b), we can write the particle flux as follow

$$P_{ny} = -\frac{n\mu}{e}\left(-e\mu B\frac{\partial V_{gr}}{\partial x} + \frac{eD_n}{n_{eq}\mu}\frac{\partial n}{\partial y}\right) \quad (s15a)$$

$$P_{py} = \frac{p\mu}{e}\left(e\mu B\frac{\partial V_{gr}}{\partial x} - \frac{eD_p}{p_{eq}\mu}\frac{\partial p}{\partial y}\right) \quad (s15b)$$

Using the Eqs. s14(a), s14(b), s15(a) and s15(b) combined with the fact the charge carrier fluxes cancel at the edges of the Hall bar we can show after some calculations that the electron and hole doping write

$$n = n_{eq} + \frac{n_{eq}\mu^2 L_n}{D_n} B \frac{\partial V_{gr}}{\partial x} \frac{sh\left(\frac{y}{L_n}\right)}{ch\left(\frac{W}{2L_n}\right)} \quad (s16a)$$

$$p = p_{eq} + \frac{p_{eq}\mu^2 L_p}{D_p} B \frac{\partial V_{gr}}{\partial x} \frac{sh\left(\frac{y}{L_p}\right)}{ch\left(\frac{W}{2L_p}\right)} \quad (s16b)$$

### III. Longitudinal profile of $V_{gr}$, $n$ and $p$ for the 1 $\mu m$ wide Hall bar

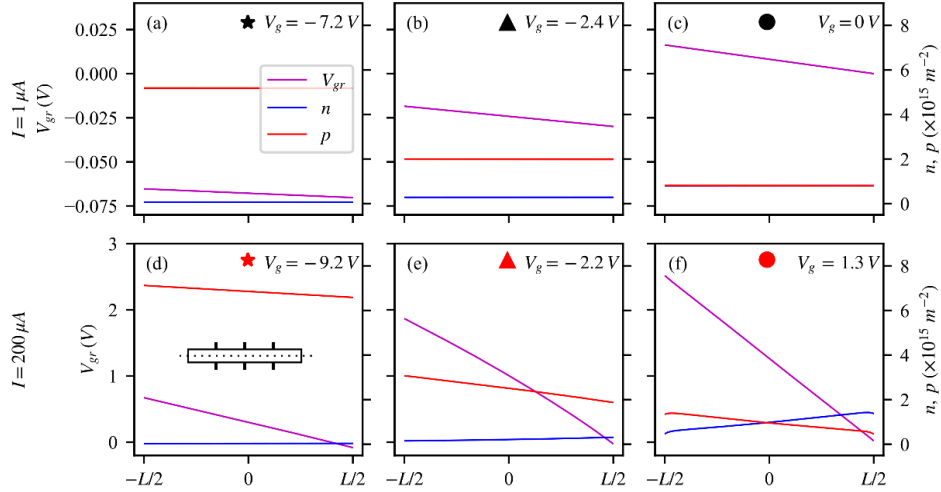

FIG. S3. Charge carrier doping $n$ (blue curves) and $p$ (red curves), electrostatic potential $V_{gr}$ (violet curves) along $x$ for $y = 0$ $\mu m$ for the 1 $\mu m$ width Hall bar at three different gate voltage values (see star, triangular and circular symbols on figure 3) and for two bias current values, 1 $\mu A$ (a), (b), (c) and 200 $\mu A$ (d), (e), (f). Insert Fig. S1(d): location of the profiles on the Hall bar.

## IV. AFM images of the GHS

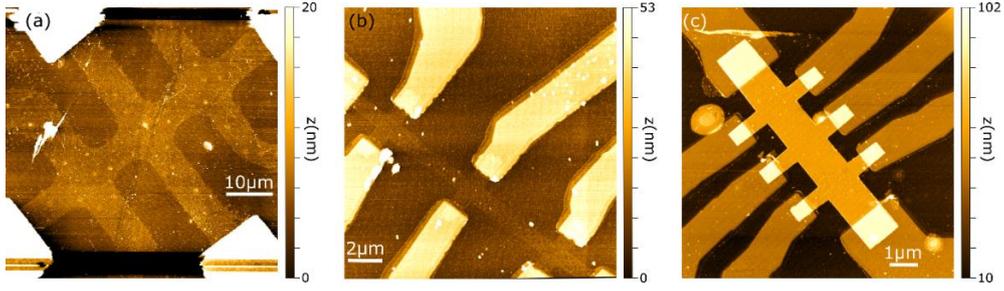

FIG. S4. (a), (b) and (c) AFM image of the CVD-GHS, HOPG-GHS and hBN-GHS respectively. The scratch observed in the middle of the image was made during the observation and after the electrical characterization. The images were obtained with a CSI Nano-Observer in resonant mode using a tip with a stiffness constant equal to $1.6\ N/m$.

## V. Joule heating modelling

Numerical simulations were performed to evaluate the temperature increase induced by Joule effect heating in the three kinds of devices. For this purpose, the heat equation in the steady state was solved numerically [1]

$$-\vec{\nabla}.\left(-\kappa_{gr} \times t_{gr} \times \vec{\nabla}T_{gr}\right) + p_J - g_{th}(T_{gr} - T_s) = 0 \tag{s17}$$

$\kappa_{gr}$ is the in-plane thermal conductivity of graphene, $t_{gr} = 0.34\ nm$ is the graphene thickness, $T_{gr}$ is the graphene temperature, $p_J$ the Joule effect heating power per unit square, $g_{th}$ is the thermal conductivity of the graphene-substrate and $T_s$ is the base temperature of the experiment. $p_J = -\vec{\nabla}V_{gr}.\vec{J}_{cond}$ was evaluated inside the graphene monolayer using the values of the graphene electrostatic potential $V_{gr}$ and the conduction current density $\vec{J}_{cond} = -(\bar{\bar{\sigma}}_n + \bar{\bar{\sigma}}_p)\vec{\nabla}V_{gr}$ determined by the numerical simulations modelling the galvanomagnetic properties of the GHS. Below the metallic contact, we estimate $p_J$, assuming the major part of the current is passing through the metallic contact, with $p_J = -\vec{\nabla}V_c.\vec{J}_{cond}$, $V_c$ being the electrostatic potential inside the contact and $\vec{J}_{cond} = -\sigma_c\vec{\nabla}V_c$, with $\sigma_c = 1\ S$ being the metal conductivity. $g_{th}$ was evaluated using the following relationships (s18a for CVD and HOPG-GHS, s18b for hBN-GHS and s18c for three samples under the metallic contact)

$$g_{th}^{-1} = R_{gr/SiO_2} + \frac{t_{SiO_2}}{\kappa_{SiO_2}} + R_{SiO_2/Si} \tag{s18a}$$

$$g_{th}^{-1} = R_{gr/hBN} + \frac{t_{hBN}}{\kappa_{hBN}} + R_{hBN/SiO_2} + \frac{t_{SiO_2}}{\kappa_{SiO_2}} + R_{SiO_2/Si} \tag{s18b}$$

$$g_{th}^{-1} = R_{metal/gr} \tag{s18c}$$

$R_{gr/SiO_2}$ is the interfacial thermal resistivity between graphene and $SiO_2$, $R_{SiO_2/Si}$, between $SiO_2$ and Si substrate, $R_{gr/hBN}$, between graphene and hBN and $R_{hBN/SiO_2}$, between hBN and $SiO_2$. $\kappa_{SiO_2}$ and $\kappa_{hBN}$ are the thermal conductivity of $SiO_2$ and hBN (out of plane) and $t_{SiO_2}$ and $t_{hBN}$ are the $SiO_2$ and hBN thickness [2]. Using values reported in the literature ($R_{gr/SiO_2} = 2 \times 10^{-8}\ K.m^2/W$ [3], $R_{gr/hBN} = 10^{-7}\ K.m^2/W$ [2], $R_{hBN/SiO_2} = 2.2 \times 10^{-7}\ K.m^2/W$ [2], $R_{SiO_2/Si} = 10^{-8}\ K.m^2/W$ [2], $\kappa_{gr} = 1000\ W/(m.K)$ [2], $\kappa_{hBN} = 3\ W/(m.K)$ [2], $\kappa_{SiO_2} = 1,3\ W/(m.K)$ [2]) and $t_{SiO_2} = 90\ nm$ for CVD-GHS, $285\ nm$ for HOPG and hBN-GHS and $t_{hBN} = 20\ nm$ for the hBN-GHS, we obtain $g_{th}$ equals to $10\ MW/(m^2.K)$ for the CVD-GHS, $4\ MW/(m^2.K)$ for the HOPG-GHS and $2.8\ MW/(m^2.K)$ for the hBN-GHS. Below the metallic contact we used $R_{metal/gr} =$ $4 \times 10^{-8} \, K.m^2/W$ [4]. It is important to note that the values of $R_{metal/gr}$ and $\kappa_{gr}$ have low impact on the results, as the majority of the Joule effect heating power is dissipated through the dielectric substrate.

FIG. S5 shows the average of the Joule effect heating power and the average of the temperature as a function of the gate voltage at high bias current for the three devices ((a) CVD-GHS, (b) HOPG-GHS and (c) hBN-GHS). We observe that the maximum temperature increase remains low, $\Delta T_{gr} = T_{gr} - T_s = 1.1 \, K$ for CVD-GHS with a maximum $p_J = 11.9 \, MW/m^2$, $\Delta T_{gr} = 6.4 \, K$ for HOPG-GHS with $p_J = 26.7 \, MW/m^2$ and $\Delta T_{gr} = 3.5 \, K$ with $p_J = 10.1 \, MW/m^2$, justifying to neglect the Joule effect heating.

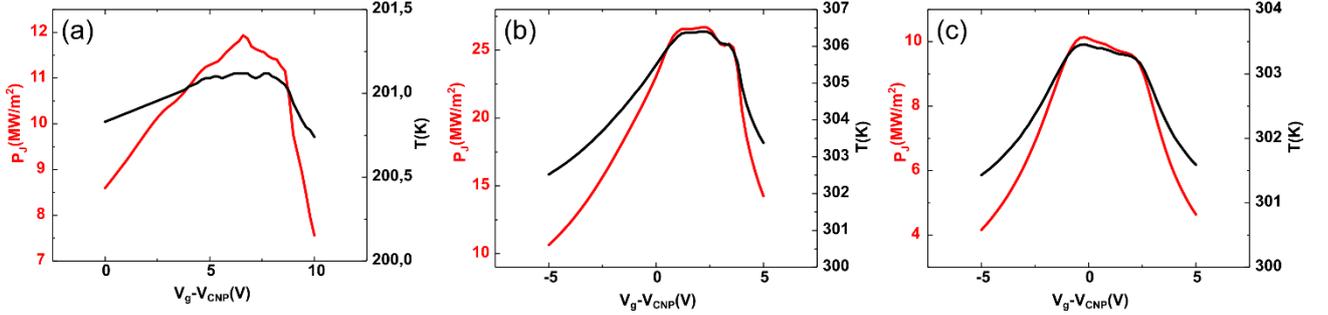

FIG. S5. Joule effect heating power, $p_J$ (red lines) and temperature $T_{gr}$ (black lines) inside the graphene monolayer for CVD-GHS (a), HOPG-GHS (b) and hBN-GHS (c) as a function of gate voltage $V_g$, $V_{CNP}$ being the gate voltage position of the CNP at low bias current.

## VI. CVD electrical characteristics

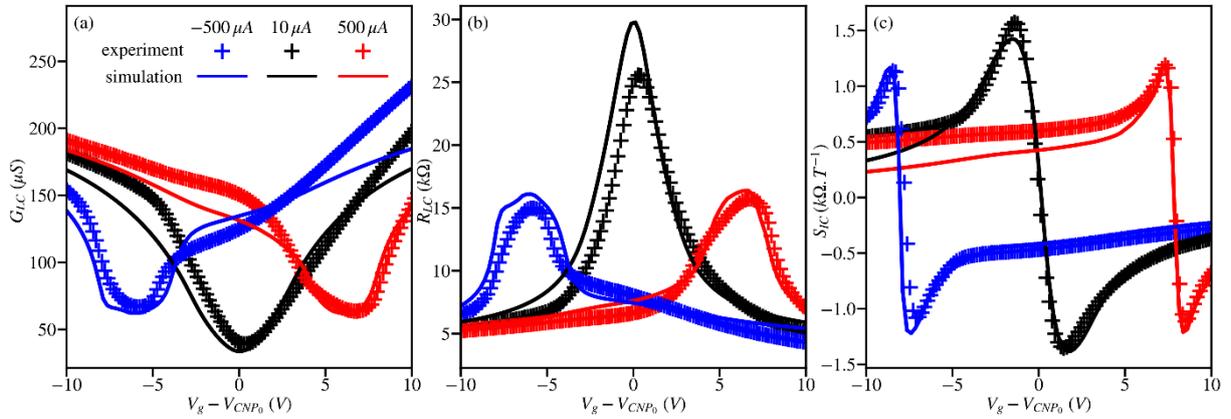

FIG. S6. Experimental (cross) and simulated (line) longitudinal conductance $G_L$ (a), resistance $R_L$ (b) and magnetic field sensitivity $S_I$ (c) of the CVD GHS at three different bias current, -500 µA (blue curves), 10 µA (black curves) and 500 µA (red curves) performed at 200 K and 1 T.

## VII. Performance degradation with width decrease

Simulations were performed on pristine (without puddles) HOPG-GHS having $W = 2 \, \mu m$, $W = 1 \, \mu m$, $W = 0.5 \, \mu m$ and $W = 0.25 \, \mu m$ at $T = 300K$ and $B = 0.1 \, T$. The other dimensions are $W_h = W$, $W' = 2W$ and $L = 8W + 3W_h$. We used for the simulations parameters comparable to what is obtained experimentally: $\mu = 2 \, m^2/(V.s)$, $v_f = 1.2 \times 10^6 \, m/s$ and $k = 10^{-4} \, m^2/(V.s)$. We used for the oxide thickness $t_{ox} = 280 \, nm$.

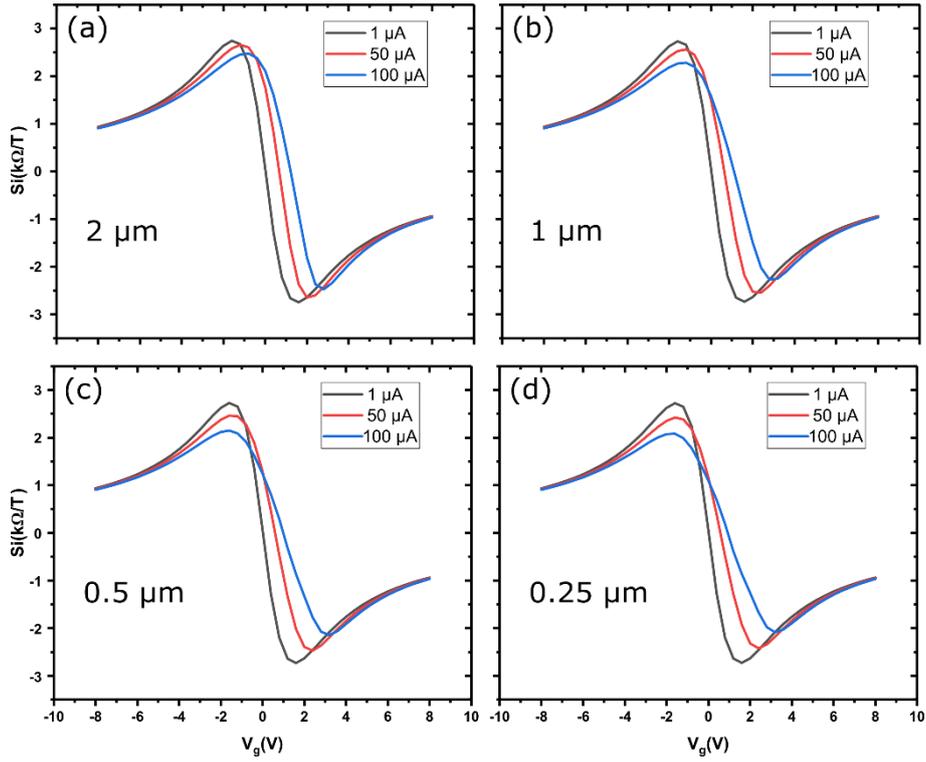

FIG. S7. (a), (b), (c) and (d) Magnetic field sensitivity of a pristine HOPG-GHS with 4 different widths.

We observe that the magnetic field sensitivity degrades (the maximum decreases and the distance separating the extrema increases) when the current increases and the width decreases. This effect is due to a more pronounced effect of the accumulation and depletion areas as the charge carrier diffusion length is $740\ nm$.